\documentclass[referee]{raa_rb}            

\usepackage{graphicx,times}             
\usepackage{natbib}
\usepackage{amssymb,amsmath}

\bibpunct{(}{)}{;}{a}{}{,}
\usepackage[pagebackref=true]{hyperref}
\usepackage{appendix}
\usepackage{enumitem} 
\usepackage{siunitx}

\usepackage{array}
\usepackage{tabularx}
\usepackage[table]{xcolor}
\usepackage{hyperref}
\definecolor{lightblue}{RGB}{184,204,228}

\begin{document}

  \title{Mock Observations for the CSST Mission: Integral Field Spectrograph--Instrument Simulation
}

   \volnopage{Vol.0 (20xx) No.0, 000--000}      
   \setcounter{page}{1}          

   \author{Zhao-Jun Yan 
      \inst{1}
   \and Jun Yin 
      \inst{1}
   \and Lei Hao 
      \inst{1}
   \and Shi-Yin Shen 
      \inst{1}
   \and Wei Chen 
      \inst{1}
   \and Shuai Feng 
      \inst{3}
   \and Yi-Fei Xiong 
      \inst{1}
   \and Chun Xu 
      \inst{1}
   \and Xin-Rong Wen 
      \inst{1}
   \and Lin Lin 
      \inst{1}
   \and Chao Liu 
      \inst{2}
   \and Lin Long 
         \inst{1}
    \and Zhen-Lei Chen 
      \inst{1}
    \and Mao-Chun Wu
      \inst{1}      
  \and Xiao-Bo Li 
        \inst{4}
    \and Zhang Ban 
          \inst{4}
    \and Xun Yang  
              \inst{4}
    \and Yu-Xi Jiang 
              \inst{4}
    \and Guo-Liang Li 
            \inst{5}
    \and Ke-Xin Li
       \inst{1} 
    \and Jian-Jun Chen 
       \inst{2}
    \and Nan Li 
       \inst{2}
    \and Cheng-Liang Wei  
            \inst{5}
    \and Lei Wang  
            \inst{5}
    \and Bai-Chuan Ren  
            \inst{6}
    \and Jun Wei 
            \inst{6}
    \and Jing Tang 
        \inst{2}
    \and Ran Li 
        \inst{7}
   }

   \institute{Shanghai Astronomical Observatory, Chinese Academy of Sciences, Shanghai 200030, China; {\it zhaojunyan@shao.ac.cn, jyin@shao.ac.cn, haol@shao.ac.cn, ssy@shao.ac.cn}\\
        \and
             National Astronomical Observatories, Chinese Academy of Sciences, Beijing 100101, China\\
        \and
             College of Physics, Hebei Key Laboratory of Photophysics Research and Application, Hebei Normal University, Shijiazhuang 050024, China.\\
        \and
             Space Optics Department, Changchun Institute of Optics, Fine Mechanics and Physics, Chinese Academy of Sciences, Changchun 130033, China.\\
        \and
             Purple Mountain Observatory, Chinese Academy of Sciences, Nanjing 210023, China.\\
        \and
             Shanghai Institute of Technical Physics, Chinese Academy of Sciences, Shanghai, 200083, China.\\
        \and
             School of Physics and Astronomy, Beijing Normal University,  Beijing 100875, China.\\
\vs\no
   {\small Received 20xx month day; accepted 20xx month day}}

\abstract{The Chinese Space Station Survey Telescope (CSST) is a next-generation Stage-IV facility renowned for its wide field of view, high image quality, and multi-band observational capabilities. Among the five instruments onboard the CSST, the Integral Field Spectrograph (IFS) offers the unique ability to simultaneously capture spatial and spectral information across a field of view of no less than $6^{''}\times6^{''}$. Key advantages of the IFS include a high spatial resolution of $0.2^{''}$ and a broad spectral coverage from 350 to \SI{1000}{\nm}, making it an ideal instrument for studying physical processes in the vicinity of supermassive black holes within galaxies. To more accurately assess the technical and scientific performance of the CSST-IFS, it is essential to develop a simulation tool that incorporates realistic effects from all optical components. Such a simulation will form an integral part of the CSST-IFS data and pipeline system, enabling the development of the data reduction pipeline well ahead of actual observations. This paper presents an end-to-end simulation workflow for the CSST-IFS, incorporating a wide range of instrumental effects that may influence its spectral and imaging performance. The simulation accounts for optical diffraction effects introduced by all components, such as image slicers and slit array, as well as sub-pixel effects from gratings. It also includes various detector noises, frame-shifting effects, and charge-transfer inefficiency. Real observational conditions—such as target Doppler shift, cosmic rays, and other in-orbit operational effects—are also considered. We describe the technical implementation of the simulation and present results that quantitatively characterize key instrument parameters.
\keywords{instrumentation: spectrographs --- techniques: imaging spectroscopy --- methods: numerical}
}

   \authorrunning{Z.-J. Yan et al}            
   \titlerunning{Mock Observations for the CSST Mission: Integral Field Spectrograph--Instrument Simulation}  

   \maketitle

%
%
\section{Introduction}           
\label{sect:intro}
The Chinese Space Station Survey Telescope (CSST) is a large astronomical space telescope of China's manned space program \citep{2011SSPMA..41.1441Z,zhan2021wide}. With a 2-meter aperture, this telescope delivers exceptional wide-field imaging and high-quality resolution, augmented by its unique capability for on-orbit maintenance and upgrades. Its primary mission is to advance research across diverse astronomical fields, including dark matter and dark energy, galaxy evolution, the Milky Way, exoplanets, and transient phenomena. These objectives will be met through advanced optical surveys and high-precision observational technologies. The telescope's main optical system gathers light, which is then analyzed by a suite of five specialized instruments: a multi-band imaging and slitless spectroscopy survey camera (SC), a multi-channel imager (MCI), an integral field spectrograph (IFS), a cool planet imaging coronagraph (CPIC), and a terahertz spectrometer (TS).

The CSST-IFS is an indispensable component of the CSST, capable of achieving spectroscopic observations on a two-dimensional area (i.e., the field of view) of specific astronomical targets. The CSST-IFS can obtain spectra over a field of view of no less than $6^{''}\times6^{''}$ and a spatial resolution of $0.2^{''}$. The wavelength coverage of CSST-IFS is 350 to \SI{1000}{\nm} with a spectral resolution of $R\ge 1000$. The CSST-IFS operates by transforming a two-dimensional field of view into spectroscopic data. Utilizing an array of 32 image slicers, it partitions the source into microscopic elements. Pupil mirrors then reassemble these elements into pseudo-slits that are dispersed onto a detector. The resulting photon data are processed and reconstructed into a three-dimensional datacube containing two spatial dimensions and one spectral dimension. This powerful output provides a complete spectrum for every spatial point and, conversely, a monochromatic image at any specific wavelength. By seamlessly combining imaging and spectroscopy, the IFS delivers simultaneous spatial and spectral resolution. As the most general-purpose instrument aboard the CSST, it enables a vast range of scientific discoveries, from the characterization of near-Earth objects to the detailed study of stellar and galactic evolution.

The performance of the CSST-IFS depends not only on its hardware but also on robust software support from a comprehensive scientific data system. This system's primary role is to generate and process raw images and spectral data into usable products for astronomers. To achieve this, a realistic observational simulation software package is crucial. This simulator will comprehensively evaluate the IFS data stream, provide essential calibration tools, and form the foundation for the data processing pipeline. Furthermore, it will drive improvements in both the IFS's hardware and software design and optimize observational strategies, ultimately ensuring the fulfillment of its core scientific objectives.

 This paper introduces the simulation philosophy of CSST-IFS, as well as its detailed implementation process and the simulation results. Section 2 outlines the comprehensive optical architecture of both the CSST and the IFS. Section 3 describes the methodological framework for optical-spectral simulations, incorporating key instrumental artifacts such as diffraction, dispersion nonlinearities, and detector quantum efficiency variations. Section 4 presents simulated data products, including calibration and observational images, as well as spectro-spatial data from a perforated plate. This section also provides direct measurements of the point and line spread functions (PSF and LSF) derived from these data. The conclusion synthesizes our findings and discusses their implications for future space-based spectroscopic instruments.



\section{The optical system introduction}
\label{sect:2}
This section details the optical design of the primary telescope and the IFS. It also describes the fundamental principles of the ground-based optical simulation developed for the telescope. A key output of this simulation is the optical path difference (OPD), which serves as essential input for the subsequent IFS software simulation and directly influences the accuracy of the spectroscopic results.
\
\begin{table}[htbp]
\centering
\renewcommand{\arraystretch}{1.5}
\setlength{\tabcolsep}{6pt}

\caption{Summary of CSST telescope and IFS}
\label{tab:csst-summary}

\begin{tabularx}{\textwidth}{|>{\centering\arraybackslash}p{3cm}|>{\centering\arraybackslash}X|}
\hline
\multicolumn{2}{|c|}{\textbf{Telescope}} \\
\hline
\multicolumn{2}{|c|}{2 m primary, off-axis TMA, focal length = 28 m, orbit $\sim$ 400 km, lifetime $\sim$ 10 years} \\
\hline
\multicolumn{2}{|c|}{\textbf{Primary Optical System / OTA}} \\
\hline
Field of view (FoV) & 1.72 deg$^2$ (full), 1.1 deg$^2$ (primary imaging) \\
\hline
Spatial res. (point/0$^\text{th}$ spec.) 
& $< 0.15^{''}/0.3^{''}$ ($\lambda = 0.6328\,\mu$m, in 1.1 deg$^2$, 80\% energy concentration region) \\
\hline
Band range & 0.255--2.5\,$\mu$m and Terahertz band \\
\hline
\multicolumn{2}{|c|}{\textbf{Integral Field Spectrograph}} \\
\hline
Observation & FoV=$6.4^{''}\times6.4^{''}$, spatial resolution $\sim 0.2^{''}$, spectral range: 0.35--1.0\,$\mu$m, $R\ge 1000$ \\
\hline
Detector
& 4k $\times$ 4k CCD (blue channel @ 0.35--0.65\,$\mu$m), 6k $\times$ 6k CCD (red channel @ 0.58--1.0\,$\mu$m) \\
\hline
Efficiency  
& Efficiency $> 40\%$ @ 0.42--0.56\,$\mu$m \& 0.62--0.82\,$\mu$m, S/N $\ge 10$ (B band 17 mag/arcsec$^2$ source with 20 $\times$ 300\,s exposure) \\
\hline
\end{tabularx}
\end{table}

\subsection{The primary optical system}

\begin{figure}
   \centering
 \includegraphics[width=\textwidth, angle=0, scale=1.0]{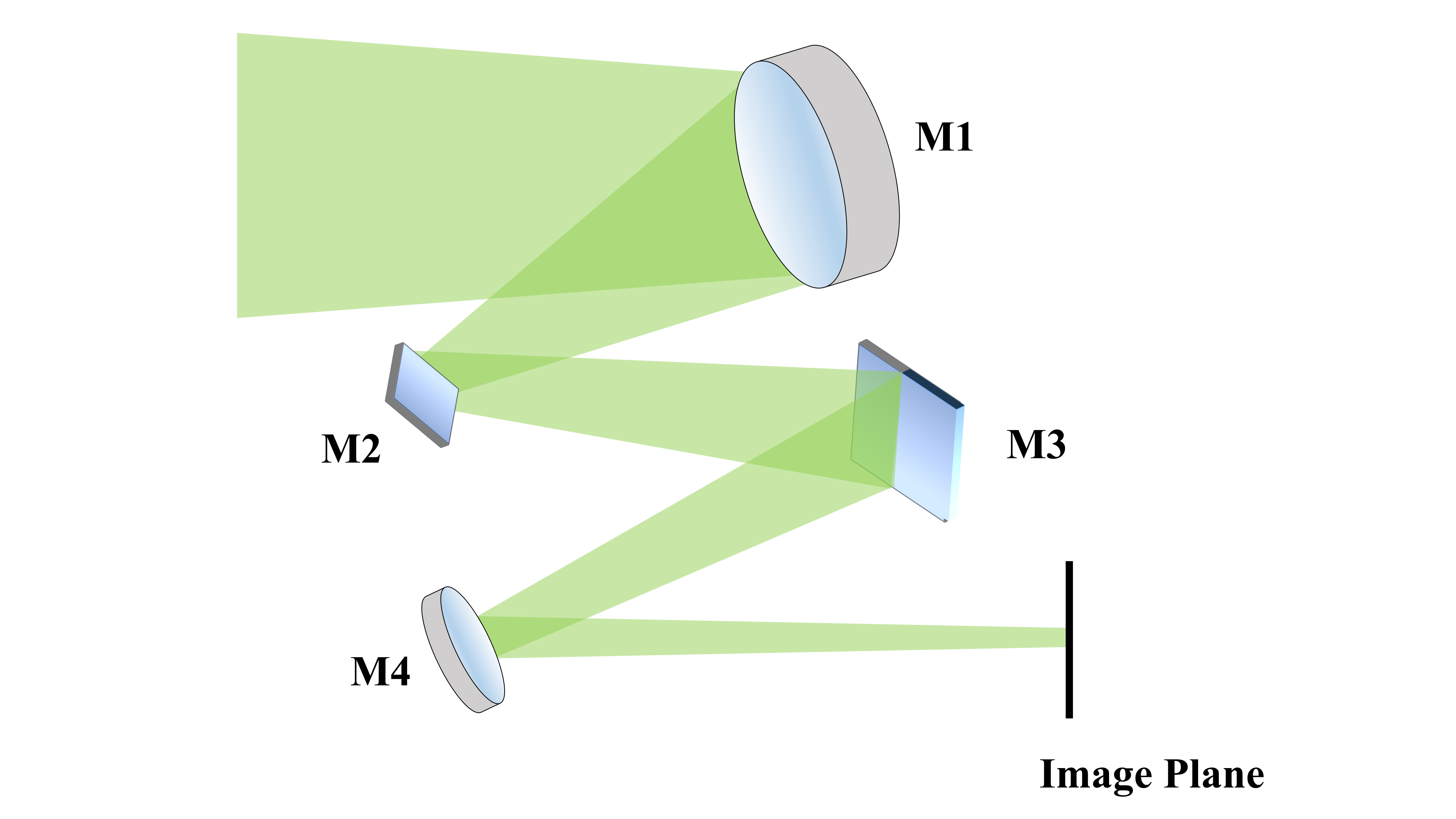}
   \caption{Schematic diagram of the optical path of the optical system of CSST.}
   \label{Fig1}
\end{figure}

The CSST main telescope system is an off-axis three-mirror anastigmat (TMA) optical system shown in Figure~\ref{Fig1}, which delivers excellent performance in terms of a large field of view and high image quality. The system is composed of four optical mirrors M1-M4, of which M1 and M2 feature quadric surfaces, and M3 employs a free-form surface. M4, known as a three-in-one plane mirror, is located near the exit pupil of the system for optical path folding. In addition, M4 is used to realize the functions of the scientific instruments switching, focusing and high-frequency precision image stabilization at the same time. As shown in Table~\ref{tab:csst-summary}, the telescope features a 2-meter clear aperture, an approximately 28-meter focal length, and a 1.72-square-degree field of view.  

\subsection{IFS optical system}

\begin{figure}
   \centering
 \includegraphics[width=\textwidth, angle=0, scale=1.0]{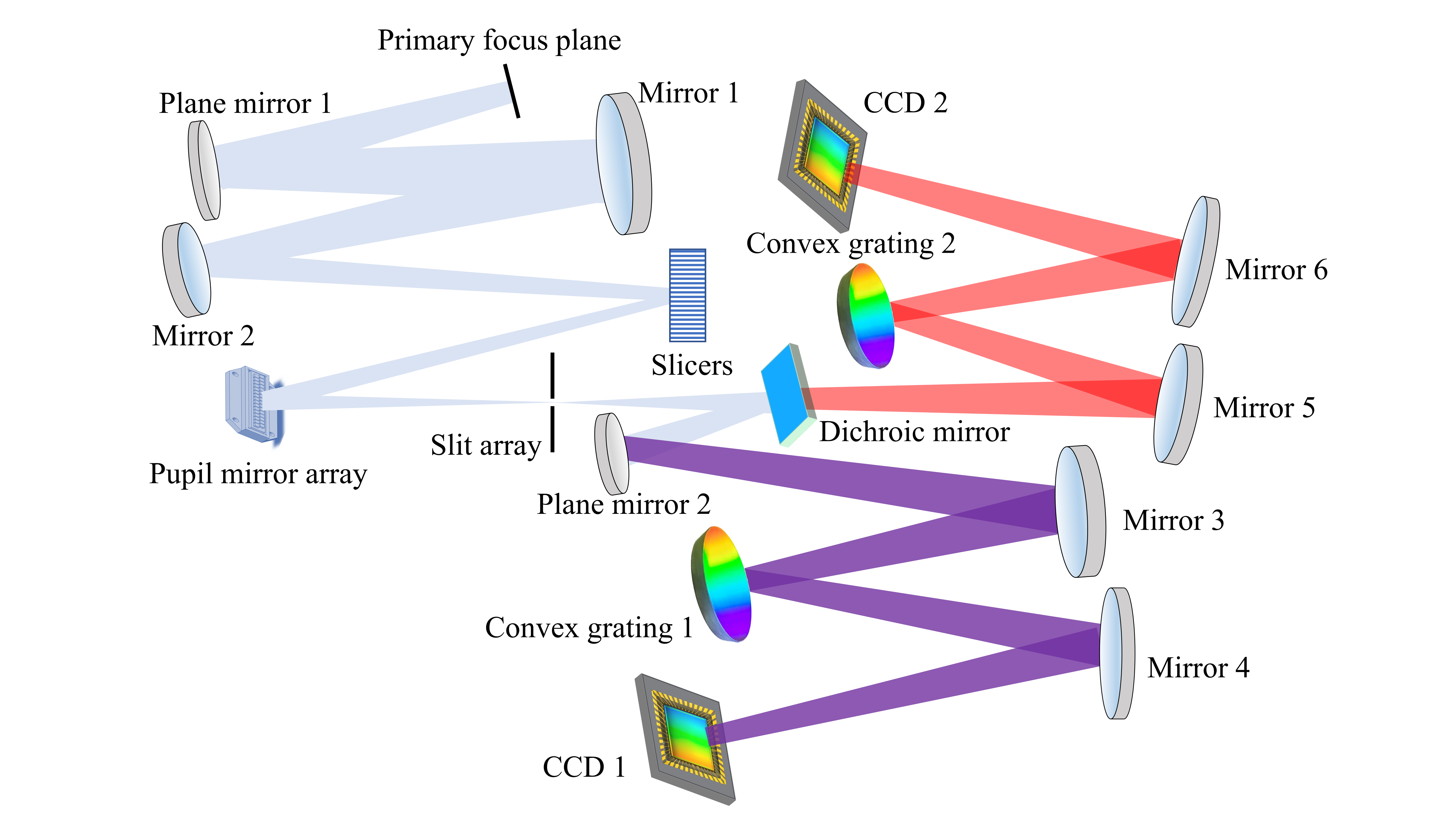}
   \caption{Schematic diagram of the optical system of IFS on CSST.}
   \label{Fig2}
\end{figure}

The IFS unit, with a field of view of $6.4^{''}\times6.4^{''}$, is installed at the focal plane of the main telescope system. It covers wavelengths from 350 to \SI{1000}{\nm} with an average spatial resolution of $0.2{''}$. These characteristics are summarized in Table~\ref{tab:csst-summary}. However, the physical size of the image slicers cannot directly match the microscopic scale of the image at the telescope's focal plane. Therefore, the image first needs to be magnified before it can be effectively sliced. The schematic diagram of the optical system of the IFS is shown in Figure~\ref{Fig2}. Plane Mirror 1 redirects the beam from the main telescope's focal plane into the IFS. Subsequently, Mirror 1 and Mirror 2 together form an image magnification system that enlarges the image by a factor of approximately 9. The magnified image is cut and reflected by the slicers. Following this, the pupil mirror array reflects and demagnifies it by a factor of 9, after which it enters the slit array. The beam is then reflected and transmitted by the dichroic mirror. The reflected beam passes through plane mirror 2 and enters the Offner spectrometer, which consists of mirror 3, convex grating 1, mirror 4 and CCD 1, covering the wavelength range of approximately 0.35-0.65 microns. The transmitted beam enters the second Offner spectrometer composed of mirror 5, convex grating 2, mirror 6 and CCD 2, with the wavelength coverage of approximately 0.58-1.0 microns.

\subsection{Engineering simulation of CSST}

A key concern is the wavefront error in the optical systems of the CSST and its IFS. To provide optical image quality input for the CSST scientific data system, the research team conducted an engineering simulation of the CSST and calculated the wavefront aberration of the in-orbit telescope under the influence of multiple error sources,  which is discussed in a companion paper (Ban et al., in preparation 2025). The simulation accounts for errors originating from optical design, mirror processing, assembly, gravitational changes, and thermal deformation. Figure~\ref{Fig1} depicts the optical path diagram of the optimally-designed telescope system. The low-frequency and mid-frequency surface shape error simulation data is combined as optical processing errors, with the final RMS degradation being 0.013 wavelengths. The CSST's entry into orbit causes changes in its gravity field, which in turn affects the position and surface shape of its mirrors. Through finite element simulation analysis, the researchers have determined the rigid body displacement and surface shape changes of each mirror upon entry into orbit. In addition, two major dynamic errors that occur during the CSST's orbital observations are considered: the high-frequency optical axis jitter error caused by the micro-vibration environment; and the low-frequency optical axis shake error caused by the residual precision image stabilization. These errors are also taken into account and simulated. CSST engineering simulations obtained the wavefront error in the IFS field of view. This wavefront error serves as one of the input parameters for IFS spectral image simulations. The significance of obtaining this error lies in the fact that these aberrations ultimately affect the spatial resolution of the IFS, thereby influencing its spectral resolution, and consequently impacting the scientific research conducted with the IFS. Therefore, the objective of CSST engineering simulations is to replicate the aberrations in the IFS field of view as accurately, objectively, and realistically as possible.

\section{IFS instrument simulation}
\label{sect:4}

\begin{figure}
   \centering
 \includegraphics[width=\textwidth, angle=0, scale=1.0]{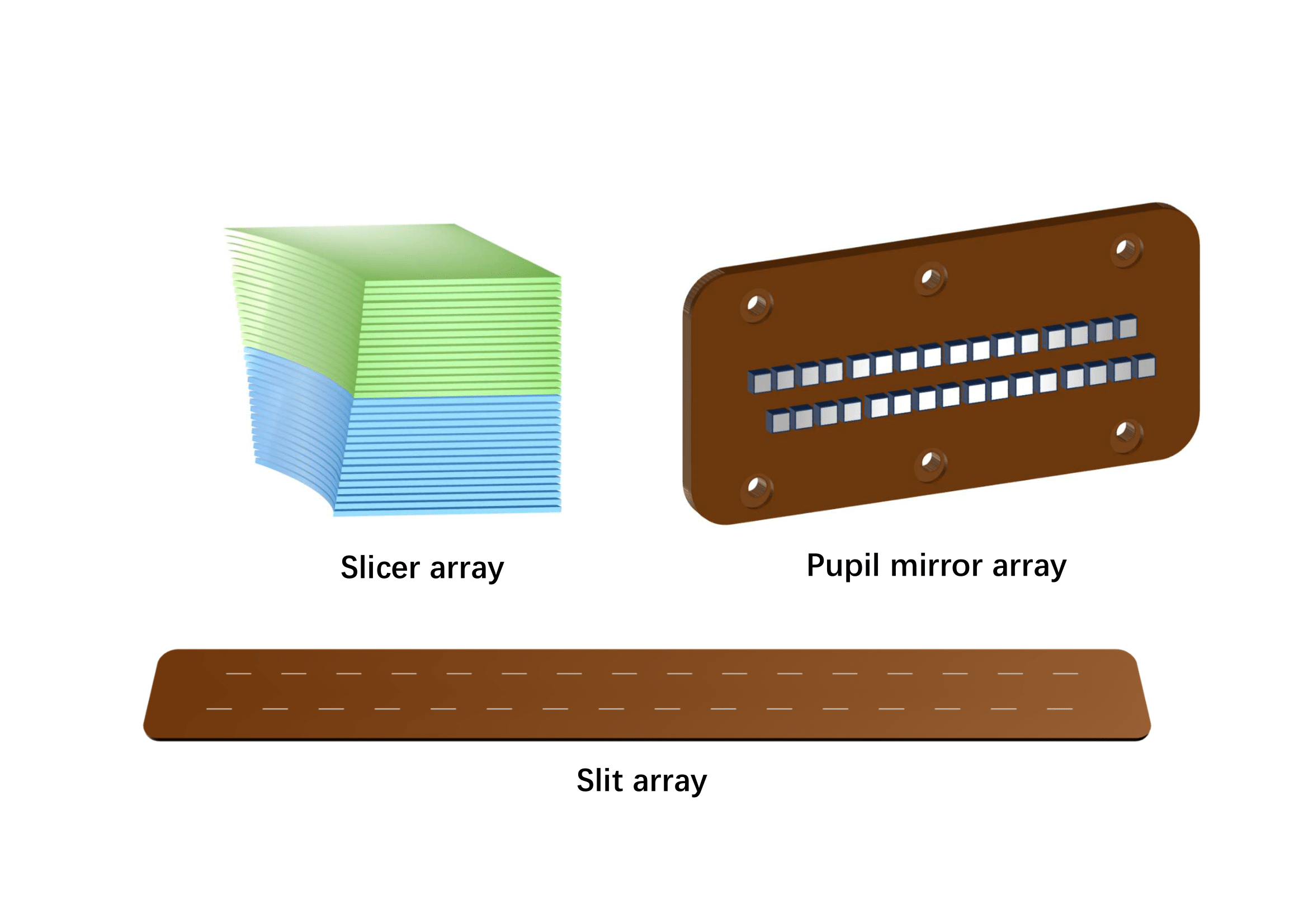}
   \caption{Schematic diagram of slicer array and pupil-mirror array.}
   \label{slicer_pupil_mirror}
\end{figure}

\begin{figure}
   \centering
 \includegraphics[width=\textwidth, angle=0, scale=1.0]{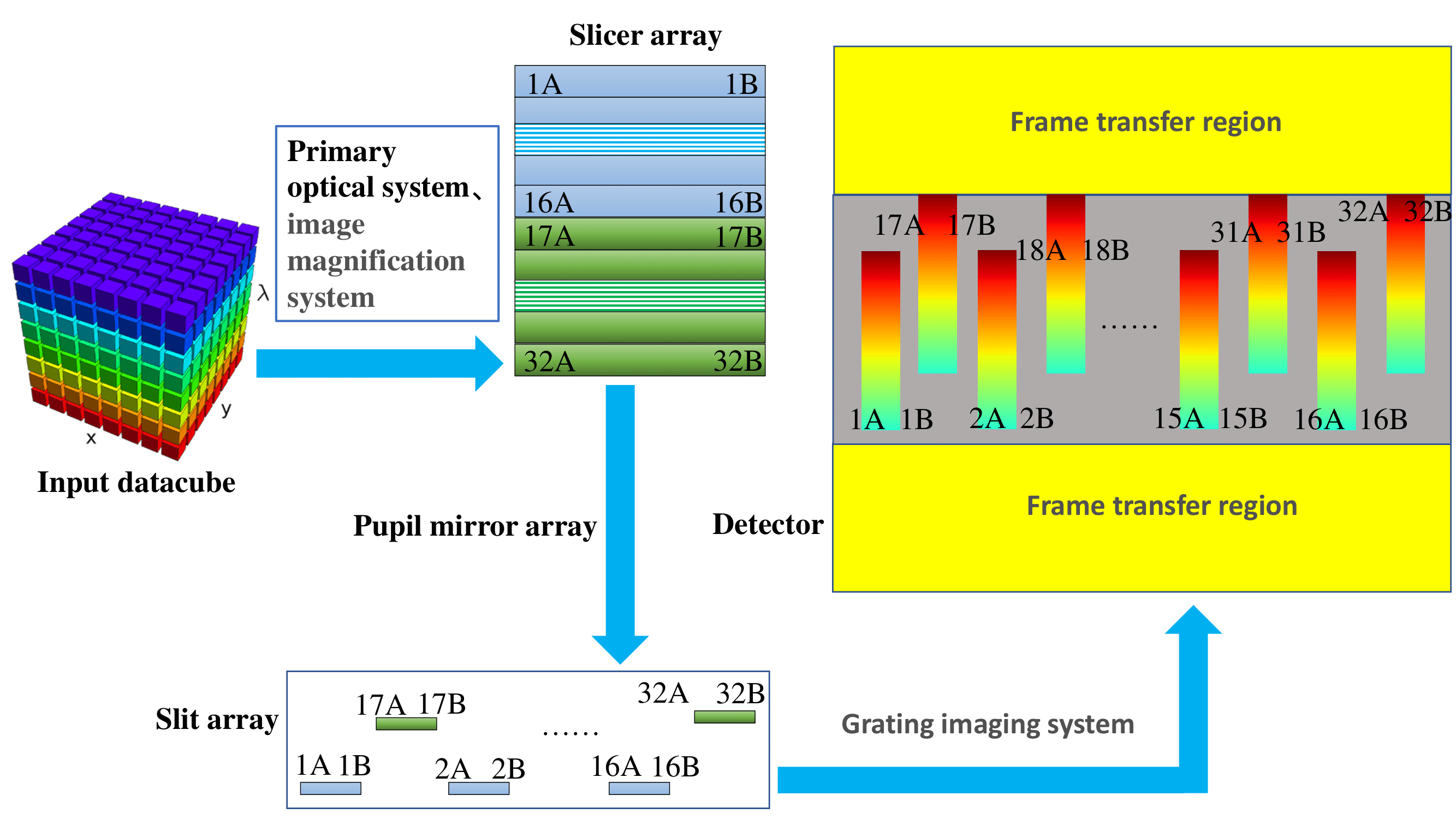}
   \caption{Flowchart of the IFS spectral imaging simulation process.}
   \label{sim_pipeline}
\end{figure}

The core task of IFS simulation is the modeling of the IFS spectral image. The main factors that affect the quality of IFS spectral imaging include the main optical system, the slicer array, the pupil mirror array, the slit array, the grating spectroscopy optical system, and the scientific camera imaging system. The general simulation workflow for IFS spectral imaging is illustrated in Figure~\ref{sim_pipeline}. The input to the simulation is three-dimensional data of the observed source, which consists of a two-dimensional spatial intensity distribution and a third dimension representing the spectral distribution. From this 3D dataset, a series of two-dimensional intensity maps—each corresponding to a specific wavelength—are sequentially extracted. Each monochromatic image is then propagated through the primary optical system and an image magnification system. Aberrations in these systems introduce optical diffraction effects during propagation, which alter the intensity distribution of the input image before it arrives at the image slicer unit, located at the system's focal plane. This degradation can be mathematically modeled as the convolution of the original image with the PSF of the optical system. The image slicer unit then segments the resulting convolved image by reflecting it from the slicing elements, thereby dividing the field into multiple sub-images. As shown in Figure~\ref{slicer_pupil_mirror}, the image slicer unit consists of upper and lower arrays, each comprising 16 slicing elements. Every element incorporates a concave reflective surface measuring 0.25 mm in thickness—equivalent to a spatial angle of 0.2 arcseconds—and 9 mm in length. The 32 resulting sub-images are then directed toward a pupil-mirror array, shown schematically in Figure~\ref{slicer_pupil_mirror}. This assembly consists of 32 individual concave spherical mirrors. The pupil-mirror array performs two critical functions: first, it spatially reconfigures the segmented sub-images to align with the slit array of the downstream spectrometer; second, it demagnifies the image scale to conform to the prescribed dimensions of the entrance slits. Following reflection by the pupil-mirror array, the 32 sub-images are coupled into the slit array, where each slit is engineered with a width of 0.03 mm and a length of 1.11 mm. These sub-images then enter the subsequent grating imaging system, where they form a series of two-dimensional spectral patterns on the detector’s imaging area. Markings such as "1A" and "1B" in Figure~\ref{sim_pipeline} serve to identify specific slicer and slit units, their orientations, and the corresponding locations on the detector where the spectral data—corresponding to signals acquired by each slicer—are recorded. By iteratively processing all wavelength-specific 2D intensity slices through this complete simulation chain, a complete spectral image of the observed target can be constructed through superposition.

\begin{figure}
   \centering
 \includegraphics[width=\textwidth, angle=0, scale=1.0]{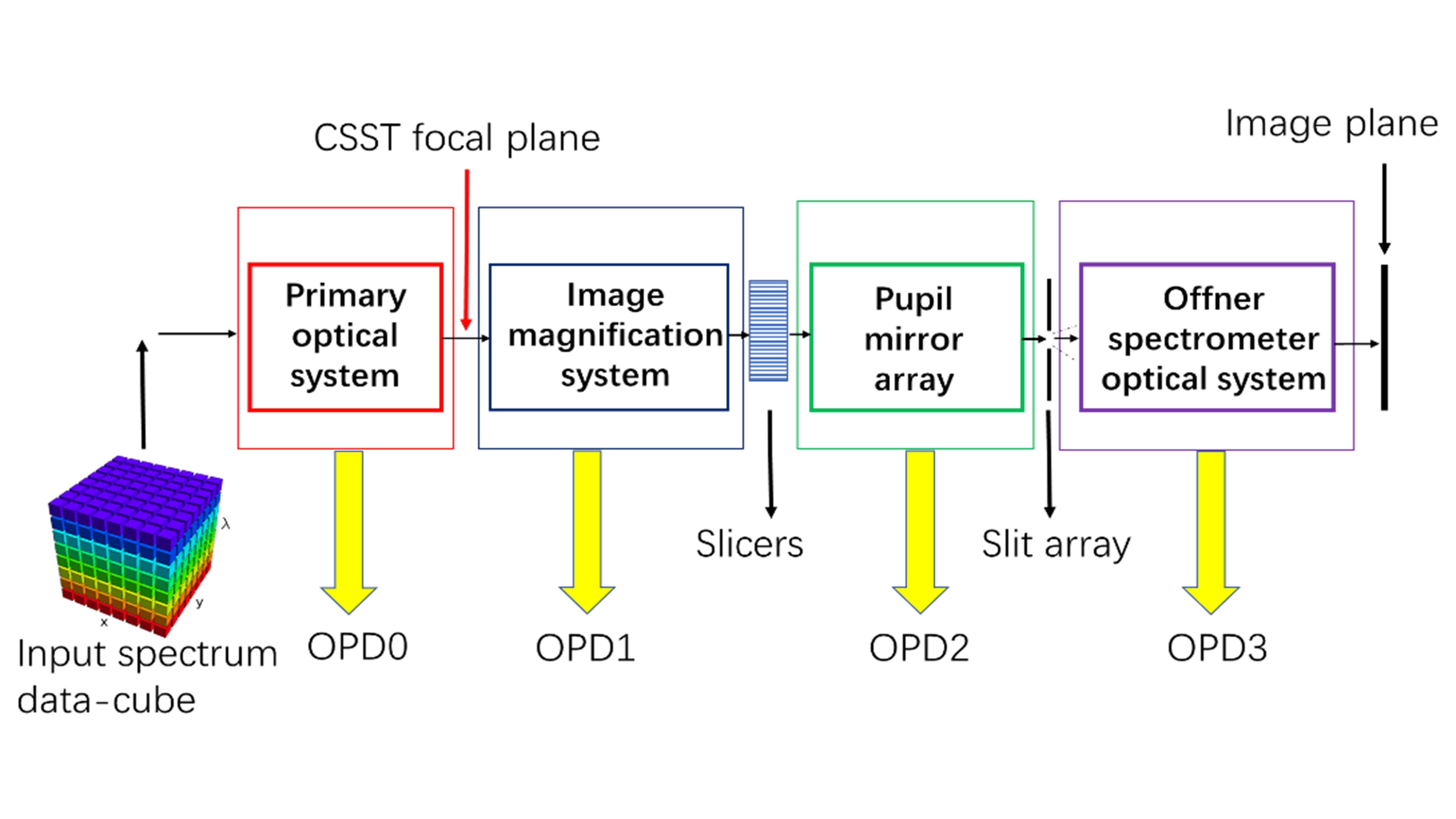}
   \caption{Representative schematic of an imaging spectrometer for IFS instrument simulation.}
   \label{Fig3}
\end{figure}

 Spectral simulation requires full consideration of the errors introduced by each part of the optics within the CSST and the IFS optical system. To accurately simulate the output spectral image in a dispersive imaging spectrometer, a physical model of the optics, as shown in Figure~\ref{Fig3}, is required to account for all components of the CSST and the IFS. The OPD of the main telescope system is represented by OPD0, while OPD1 represents the OPD of the IFS image magnification system. Similarly, OPD2 represents the OPD of the pupil mirror system, and OPD3 represents the OPD of the grating spectrometer system. This model enables the simulation of the impact of aberrations in individual optical components and assemblies on the spatial diffraction characteristics of the source spectrum. Since the slicer unit and slit array are located at the image plane, where the system operates under diffraction-limited conditions, the model can accurately simulate how these components modulate the amplitude of the pupil function during beam propagation.

The input spectral data are divided into two types: (1) the simulated spectral data of the observation target generated by the Gehong module (which is described in a companion paper: Feng et al., in preparation), and (2) the calibration lamp spectral data used for instrument calibration. The Gehong module creates an astronomical datacube to simulate the spectrum of the observed source. The datacube here is defined as a collection of objects placed within the field of view of the instrument, with spatial and spectral properties that can be mapped to the IFS detectors.  The Gehong module requires as input the two-dimensional spatial distributions of various galactic physical parameters. These include the parameters for generating the stellar continuum (e.g., magnitude, star formation history, chemical enrichment history ), those for modeling the ionized gas emission lines (e.g., H$\alpha$ flux, gas-phase metallicity \citep{Sanchez2014}), as well as the kinematic properties of galaxies \citep{Cappellari2006,Genzel2011}, namely the line-of-sight velocity and velocity dispersion, and the color excess for modeling the effect of dust extinction. Based on these input parameters, the module finally generates a three-dimensional datacube of the galaxies or stars. The code for the Gehong module can be downloaded at the following URL: \url{https://csst-tb.bao.ac.cn/code/csst-sims/csst_ifs_gehong}.

The spectral datacube is a three-dimensional matrix containing the two-dimensional spatial information and the one-dimensional spectral information of the observed target. We assume that the input spectral datacube is represented as a three-dimensional matrix \(\mathcal{A} \in \mathbb{R}^{M \times N \times P}\). A matrix slicing operation yields the two-dimensional intensity distribution \(Img_0(\lambda) = \mathcal{A}(:,:,l)\), where \(\lambda = \text{Wave}(l)\) denotes the corresponding wavelength at spectral index \(l\). The wavelength interval between adjacent slices determines the spectral resolution of the input spectrum. To balance computational efficiency and spectral fidelity, a slice interval of \SI{0.1}{\nm} is adopted across the input spectrum, which spans from \SI{300}{\nm} to \SI{1100}{\nm}.  The image $Img_{0}(\lambda)$ passes through the CSST primary optical system and then the resulting image $Img_{1}(\lambda)$ in the focal plane of the CSST is calculated as 
\begin{center}
\begin{equation}\label{eq 1}
Img_{1}(\lambda)=Img_{0}(\lambda) \ast PSF_{0}(\lambda)
\end{equation}
\end{center}
where $PSF_{0}(\lambda)$, calculated from the OPD0 data, denotes the PSF of the CSST primary optical system at the wavelength of $\lambda$ and $\ast$ denotes convolution operation. The spatial variation of the PSF is justifiably neglected in our simulations due to the exceptionally small field of view of the IFS. In the image plane where the slicer array is located, the image $Img^{'}_{1}(\lambda)$ which is not reflected yet by the slicer array, is calculated as 
\begin{center}
\begin{equation}\label{eq 2}
Img^{'}_{1}(\lambda)=Img_{1}(\lambda) \ast PSF_{1}(\lambda)
\end{equation}
\end{center}
where $PSF_{1}(\lambda)$, calculated from the OPD1 data, denotes the PSF of the image magnification system. The image $Img^{'}_{1}(\lambda)$ is then sliced by the slicer array, which slices the image into 32 separate sub-images, corresponding to the 32 slicers in the array. For ease of calculation, the sub-images have the same matrix dimensions as the image $Img_{0}(\lambda)$. The sliced sub-image $Img_2(k, \lambda)$ is calculated as
\begin{center}
\begin{equation}\label{eq 3}
Img_2(k,\lambda)=Img^{'}_{1}(\lambda) \times Mask\_slicer(k)
\end{equation}
\end{center}
where $Mask\_slicer(k)$ is a mask matrix in which the element value is 0 or 1 and has the same matrix dimensions as $Img^{'}_{1}(\lambda)$. This means, for example, that when $k=1$, only the elements of the image $Img_{2}(1, \lambda)$ that correspond to the portion of the field reflected by the first slicer have non-zero values, while the rest of the elements of the image $Img_{2}(1, \lambda)$ are zero. The sub-image $Img_{2}(k, \lambda)$ is received and reflected by the pupil mirror array one by one, and then passes through the slit array. The slit size is so small, in the order of microns, that the diffraction effect of the slit needs to be considered. The sub-images after the slit array are obtained and calculated as
\begin{center}
\begin{equation}\label{eq 4}
Img_3(k,\lambda)=Img_2(k,\lambda) \ast PSF_2(\lambda) \times Mask\_slit(k)
\end{equation}
\end{center}
where $PSF_2(\lambda)$, calculated from the OPD2 data, denotes the PSF of the pupil mirror demagnification system and $Mask\_slit(k)$ denotes the mask function of the $k\text{-th}$ slit. $Img_3(k, \lambda)$ then passes through the Offner spectrometer, and its spatial distribution is influenced by the system's PSF, as well as being dispersed in the wavelength direction by the grating. In the spatial dimension, the new image $Img_4(k, \lambda)$ is computed as 
\begin{center}
\begin{equation}\label{eq 5}
Img_4(k,\lambda)=Img_3(k,\lambda) \ast PSF_3(\lambda) \cdot Q(\lambda)
\end{equation}
\end{center}
where $PSF_3(\lambda)$, calculated from the OPD3 data, denotes the PSF of the Offner spectrometer system and $Q(\lambda)$ represents the total system efficiency, incorporating the combined response of the entire optical systems (CSST and IFS) as well as the quantum efficiency of the detector.  Next, the effect of the grating dispersion system on the image is addressed. The grating is one of the most significant elements of the spectrometer, and it is composed of a microscopic and periodic groove structure. This structure operates by angularly dispersing incident light via diffraction, directing each constituent wavelength to propagate in a unique direction.

The effect of aberrations on the convex grating used in the IFS system has been simulated in Eq.~\ref{eq 5}. The following step is to simulate the dispersion effect of this grating on the beam. For one wavelength, there are 32 sub-images. Each sub-image $Img_4(k, \lambda) (k=1, \cdots, 32)$ is projected onto the detector, and its position on the detector depends on the relative position of the slicer to the detector, as well as the wavelength. The pixel size of the sub-image is equivalent to the actual pixel size of the detector. Here, a pixel width of one represents $0.1^{''}$. However, the offset of the beam after dispersion is not necessarily an integer multiple of the pixel, and there is usually a sub-pixel offset. 

\begin{figure}
   \centering
 \includegraphics[width=\textwidth, angle=0, scale=1.0]{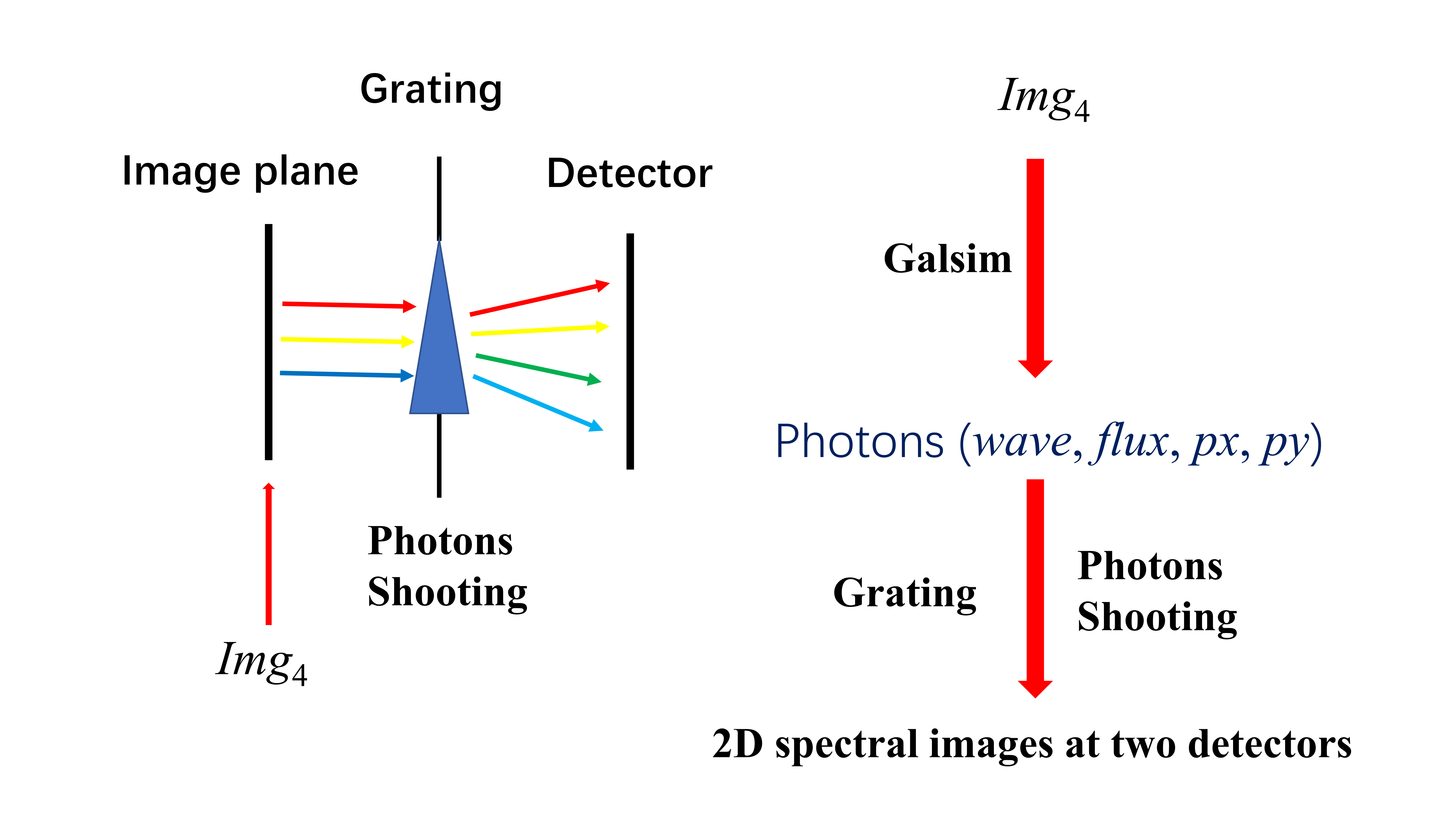}
   \caption{Schematic diagram of an image converted to photons.}
   \label{Fig4}
\end{figure}

To accurately simulate the sub-pixel effect in IFS, the "Photon shooting" module in GalSim is used to convert the image into photons, as shown in Figure~\ref{Fig4}. The sub-image $Img_4(k, \lambda) (k=1, \cdots, 32)$ is converted into a photon array, denoted as Photons($wave, flux, px, py$), where each photon is characterized by its wavelength ($wave$), flux ($flux$), and spatial coordinates ($px, py$).  

Next, each photon is projected onto the detector, and its position on the detector is determined by the formula below.
\begin{center}
\begin{equation}\label{eq 6}
\begin{cases} 
px^{'}=px+\sum_{i=0}^{5}a_i\lambda_i+Slicer_X(k)\\
py^{'}=py+\sum_{i=0}^{5}b_i\lambda_i+Slicer_Y(k)\\  
\end{cases}
\end{equation}
\end{center}
where $(px, py)$ and $(px^{'}, py^{'})$ are the coordinates before and after the photon projection onto the detector, respectively. $a_i$ and $b_i$ denote the dispersion coefficients of the grating, and we use a fifth-order polynomial to simulate the dispersion effect. $Slicer_X(k)$ and $Slicer_Y(k)$ represent the central coordinates of the original sub-image $Img_4(k, \lambda)$, that is, they represent the position of the different slicers. Finally, all the photons are received by the detector to form a spectral image which is achieved via the sensor module in GalSim. Thus far, the simulation has been conducted for only a single wavelength slice within the three-dimensional spectral data of the observed source, as processed through the CSST-IFS system. To construct a complete synthetic spectral representation of the source on the detector, the same simulation procedure must be iteratively applied to each remaining wavelength slice. This iterative process will ultimately yield a fully synthesized spectral image that replicates the observational data acquired by the detector.

The IFS employs two spectrometers. To process the simulation, the comprehensive efficiency curves of the ultraviolet and near-infrared channels are first applied in the calculation defined by Eq.~\ref{eq 5}. The subsequent steps are then carried out separately for each channel to produce the final spectral images. In the following discussion, we adopt the terminology of "blue channel" and "red channel" for the ultraviolet and near-infrared channels, respectively, owing to the predominance of their corresponding optical bands. This convention enhances both clarity and conciseness.

\subsection{PSF calculations}
A PSF is the diffraction pattern caused by light from a point source passing through an optical system, with the various optical components and obstructions adding structure to it. Given the very small field of view of the IFS, the simulation assumes that the OPD is constant across the field. The PSF calculation starts from the pupil function defined by  Eq.~\ref{eq 7} which includes the aperture function $A$ and the OPD function. 
\begin{center}
\begin{equation}\label{eq 7}
P=Ae^{(i \cdot 2\pi OPD/\lambda)}
\end{equation}
\end{center}
Fourier transform of $P$ yields the amplitude spread function (ASF) as shown in Eq.~\ref{eq 8}. 
\begin{center}
\begin{equation}\label{eq 8}
ASF=FFT(P)
\end{equation}
\end{center}
Finally, squaring the modulus of the ASF gives the PSF. However, the FFT-based implementation imposes constraints for short wavelengths, requiring zero-padding on the pupil plane and cropping operations on the focal plane. The uniform sampling interval on the two planes cannot accurately compute an under-sampled PSF for short wavelengths. A more flexible matrix triple product (MTP) method provides flexibility for PSF calculations. The discrete calculation of Eq.~\ref{eq 8} is implemented using a 2D FFT algorithm. It can also be expressed as an MTP form \citep{jurling2018techniques}
\begin{center}
\begin{equation}\label{eq 9}
ASF=\Omega_yP\Omega_x
\end{equation}
\end{center}
where $\Omega_x=e^{-i2\pi K_x^TX}$ and $\Omega_y=e^{-i2\pi K_y^TY}$. $K_x$, $K_y$, $X$ and $Y$ are the coordinates, represented by row vectors, in the spatial frequency domain and the spatial domain, respectively. $T$ represents the transpose.

As shown in Figure~\ref{Fig3}, the complete optical system of the CSST, which incorporates the IFS, is partitioned into four sections. The wavefront of the first section, termed OPD0, is derived from the engineering simulation of the primary telescope. The wavefronts of the remaining three sections, termed OPD1, OPD2 and OPD3, are generated by the random distribution of aberrations with root-mean-square (RMS) values of 0.074, 0.065 and 0.05, respectively, at the wavelength of \SI{632.8}{\nm}.

\subsection{Comprehensive efficiency of the IFS}
\begin{figure}
   \centering
 \includegraphics[width=\textwidth, angle=0, scale=1.0]{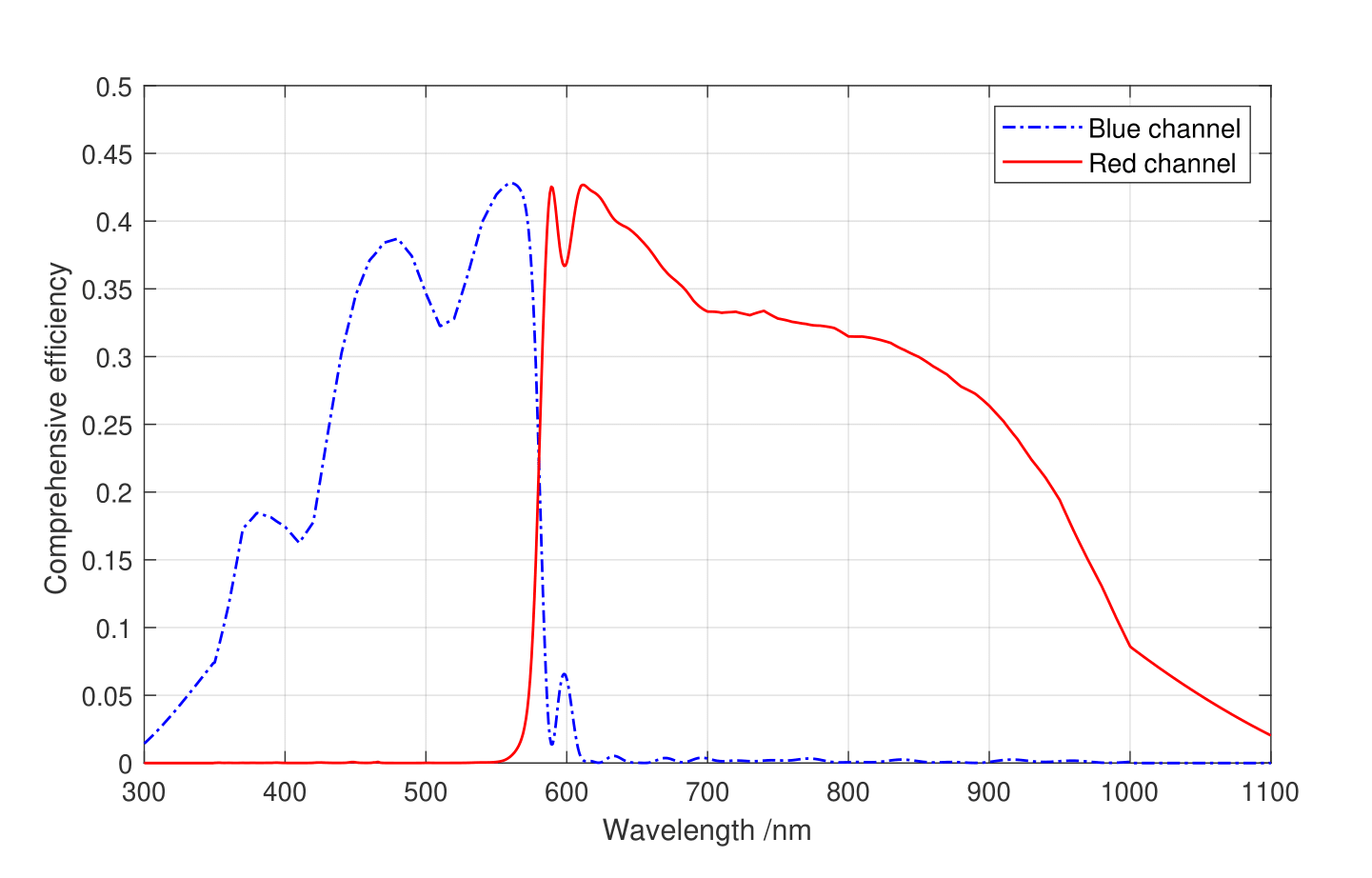}
   \caption{Efficiency curves for two channels of IFS.}
   \label{Fig5}
\end{figure}

As mentioned in Eq.~\ref{eq 5}, $Q(\lambda)$ represents the total system efficiency, incorporating the combined response of the entire optical systems (CSST and IFS) as well as the quantum efficiency of the detector. Its curves in blue and red channels are shown in Figure~\ref{Fig5}. As shown in Figure~\ref{Fig2}, the IFS employs a dichroic mirror to split the incoming broadband light by wavelength. Shorter wavelengths are almost entirely directed into the blue channel, while longer wavelengths are predominantly directed into the red channel. However, light within the intermediate wavelength range is split between both channels at certain ratios. This is why the combined efficiency curve described in the figure exhibits such distinct boundaries at the blue and red channels. As for the valleys observed in the curves, these are primarily due to the optical design of the multilayer coatings on the dichroic mirror. 

\subsection{Doppler effect}
IFS instrument simulations also take into account the Doppler effect. The simulation of the Doppler effect is based on the manipulation of the wavelength properties of photons. First, the orbit information parameters of the telescope at the start of the exposure are obtained, and the radial velocity $v_1$ of the telescope relative to the observed target is calculated. Then the shifted  wavelength of the photons due to the Doppler effect is calculated according to the following formula. 
\begin{center}
\begin{equation}\label{eq 10}
\lambda_1=\lambda_0(1+v_1/c)
\end{equation}
\end{center}
Here, $\lambda_0$ refers to the original wavelength of the photons, and $c$ is the speed of light in vacuum. It should be noted that the photons are obtained from the sub-image $Img_4(k, \lambda) (k=1, \cdots, 32)$ and they have the same original wavelength $\lambda_0$. At the end of the exposure, the radial velocity $v_2$ of the telescope relative to the observed target is recalculated, and the Doppler-shifted wavelength of the photon is calculated according to the following formula.
\begin{center}
\begin{equation}\label{eq 11}
\lambda_2=\lambda_0(1+v_2/c)
\end{equation}
\end{center}
Since the Doppler effect occurs  throughout the entire exposure, the wavelength of the photons from the image $Img_4(k, \lambda_0)$ that finally arrive at the detector follows an approximately uniform distribution over the interval [$\lambda_1, \lambda_2$].  Therefore, to simulate this effect, we replace the original wavelength value $\lambda_0$ of each photon with a new value sampled from this uniform distribution [$\lambda_1, \lambda_2$] before projecting them onto the detector. In addition to altering wavelengths, the Doppler effect also influences the spectral flux distribution. For individual photons, a change in wavelength results in a corresponding change in energy. When the wavelength shift is very small, the associated energy change is negligible. In our prior analysis of the Doppler effect, we determined that for an exposure time of 300 seconds, the wavelength shift induced by the Doppler effect is extremely small—significantly smaller than the spectral resolution of the IFS. Therefore, in our simulation software, we did not account for photon energy changes due to the Doppler effect. In other words, the total number of photons acquired over the exposure interval remains invariant under Doppler correction. However, due to the Doppler-induced shift in photon wavelengths, the observed spectral energy distribution of the target is correspondingly modified.

\subsection{Cosmic rays}
The cosmic ray simulation is based on empirical data recorded by the Hubble Space Telescope. These data include information on the length and energy of the cosmic rays. The total number of cosmic rays in each detector is a configurable input parameter for the simulation. In the simulation, the cosmic ray coverage is set to 0.5 percent for an exposure time of 300 seconds. Here, "cosmic ray coverage" refers to the ratio of the number of detector pixels affected by cosmic rays to the total number of pixels. Cosmic ray particles strike the detector at random positions and with random incidence angles. When a particle strikes the detector, it produces a cosmic ray trace. The energy distribution of these cosmic rays follows a power law, and each event is characterized by its length and energy, which are drawn from corresponding distributions \citep{5205506}. Notably, the cosmic ray coverage value of 0.5\% was chosen primarily to meet the requirements of subsequent data processing pipelines, enabling validation of the cosmic ray rejection algorithm’s effectiveness. The software allows this coverage ratio to be arbitrarily adjusted according to specific needs.

\subsection{Backgrounds and stray light}
The brightness of the background emission directly determines the achievable depth of many IFS observations. There are three main components of the background; namely zodiacal light, earthshine, and scattered light. The magnitude of the zodiacal light is influenced by the angle of the target relative to the Sun and the ecliptic. Data from Leinert \citep{leinert19981997} are used for zodiacal light simulation. Earthshine varies highly as a function of the angle between the target and the Earth limb. Data from the Figure 9.1 in the HST/WFC3 instrument handbook (\url{https://hst-docs.stsci.edu/wfc3ihb/chapter-9-wfc3-exposure-time-calculation/9-7-sky-background, https://cads.iiap.res.in/tools/zodiacalCalc/Documentation}) are used in our earthshine simulations. Stray light is generally used to describe unwanted light that enters the focal plane of an optical system and is a critical factor that has a significant impact on high precision photometry of optical astronomical telescopes, as it has the potential to reduce the signal-to-noise ratio of astronomical objects. Stray light analysis and simulation are conducted by the CSST Scientific Data Systems team. This work begins with 3D modeling of the CSST's optical and mechanical structure, followed by the assignment of surface properties and ray-tracing simulations. 

\subsection{Photo response non-uniformity effect of detector}
Photo-response non-uniformity (PRNU) represents the uniformity of a camera’s response to light, which is particularly important in high-precision photometric applications. It is defined as the standard deviation of the pixel gain values, expressed as a percentage. In the simulation, a two-dimensional Gaussian distribution is used to simulate the non-uniformity of the normalized response between pixels, with a mean of 1 and a standard deviation of 0.001.

\subsection{Charge transfer inefﬁciency effect of detector}
The high flux of damaging radiation at the altitude of the space telescope orbit generates an ever-increasing population of charge traps in the silicon of charge-coupled devices (CCDs), lowering the latter's charge transfer efficiency (CTE), which is quantified by the fraction of charge successfully transferred (clocked) between adjacent pixels. The term charge transfer inefficiency (CTI) is more useful in many cases. The principal observable consequence of CTI is that a star whose induced charge must travel across multiple pixels before being read out will appear fainter than the identical star located closer to the read-out amplifier. This effect is significant for all CCD detectors used in CSST instruments. The main cause of CTI tailing is the capture of charge by electronic traps in the pixels during the readout process. For the CTI effect added to the IFS, please refer to the trailing profile of the CTI effect measured in the reference paper \citep{anderson2010empirical}. In the current simulation, the CTI trailing direction is set along the column direction. The module, however, allows the direction to be set to either row or column. At present, the tail length is generally several pixels, and the  fraction of total charge contained in the tail is about 0.003$\%$. This corresponds to a simulated CTE of about 99.997$\%$ for the IFS detector models (E2V CCD 231-C6 and CCD 230-84). Notably, the $99.997\%$ CTE value is a conservative estimate for algorithm validation, not based on radiation tests. This value was intentionally chosen to introduce a noticeable pixel trailing effect, which is beneficial for the development and validation of CTI correction algorithms. 

\subsection{Nonlinearity effect of detector}
A CCD consists of an array of pixels arranged on a very thin silicon layer. Each incident photon is converted into an electron-hole pair. Up to 50000 to 500000 electrons can be stored per pixel, depending on the CCD. After an exposure, a controller can shift the electrons around and read them out, converting the electron charge of each pixel into digital counts (analog-to-digital units or ADU). The term gain refers to the conversion factor (electrons/ADU). Contrary to popular belief, CCDs are not perfectly linear systems. The number of ADU is not exactly proportional to the number of incident photons after bias subtraction. The nonlinearity of the controller gain is most likely due to its non-constant relationship with the number of electrons. Much larger nonlinearities occur as the CCD nears saturation because the charge collection efficiency drops in pixels whose potential wells are nearly full. The nonlinearity of the detector is simulated by the following formula.
\begin{center}
\begin{equation}\label{eq 12}
f(x)=x-a \cdot x^2
\end{equation}
\end{center}
where $x$ represents the photon counts received by the pixel, $a$ is the nonlinearity coefficient with the default value of $10^{-7}$ and $f(x)$ represents the simulated counts for this pixel affected by the detector's nonlinearity. The default value of coefficient $a$ is used primarily for testing the data processing pipeline algorithm and will be replaced with values measured from the actual detector in the future.

\subsection{Frame transfer effect of detector}
\begin{figure}
   \centering
 \includegraphics[width=\textwidth, angle=0, scale=1.0]{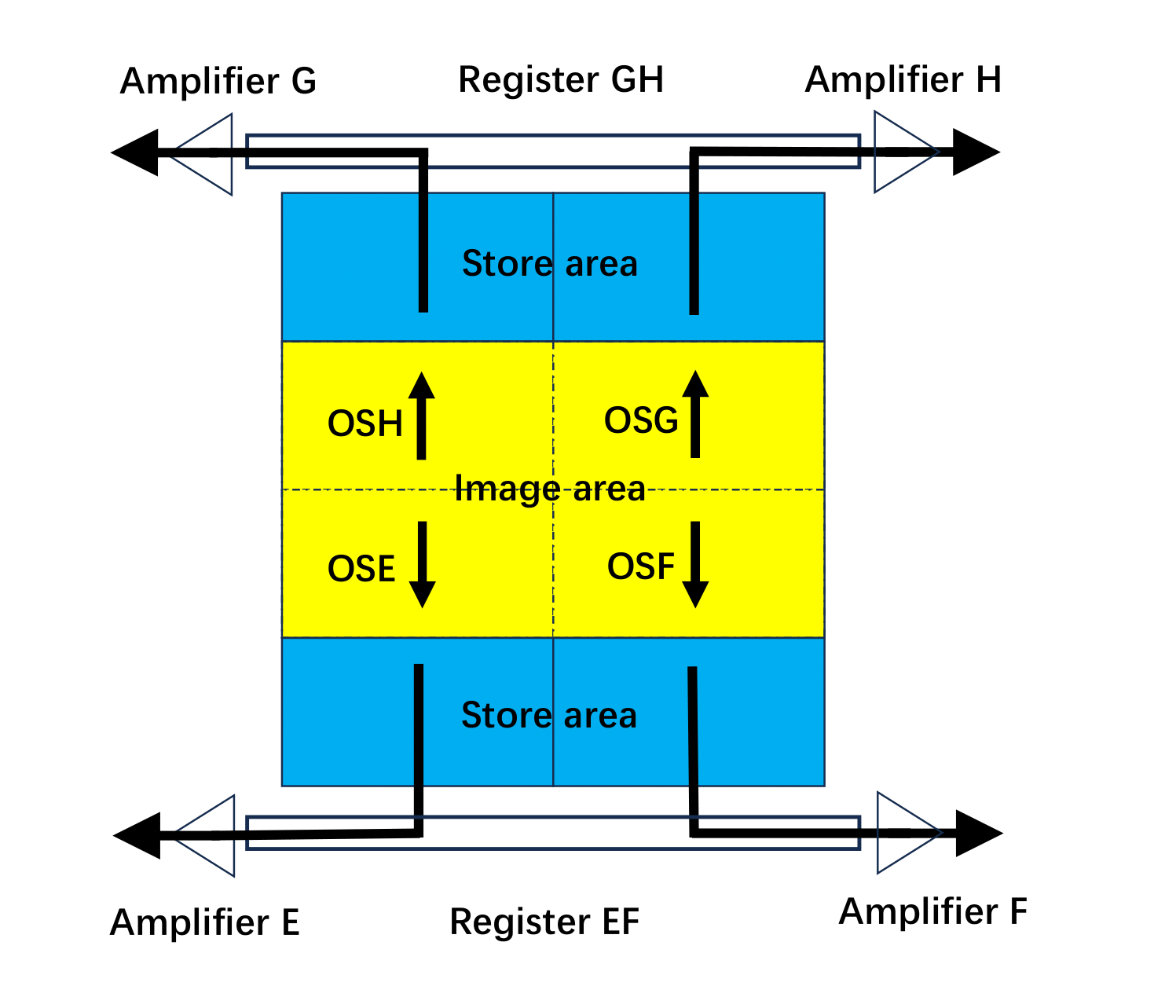}
   \caption{Schematic of chip layout.}
   \label{chip_figure}
\end{figure}

As the IFS detector system operates without a mechanical shutter, it employs two frame-transfer CCDs—the E2V CCD231-C6 and CCD230-84. These are scientific-grade image sensors known for their advanced architecture. Their "split frame transfer via four amplifiers" design enables extremely high readout speeds and low noise by combining rapid charge transfer with parallel output. To understand the operational process, it is essential to first examine the physical layout of the chip. As shown in Figure~\ref{chip_figure}, the central image area serves as the active region for exposure and photon collection. Adjacent to it, a light-shielded storage area—split into sections above and below the image region—facilitates the split frame transfer process. Charge accumulated in the image area is rapidly transferred vertically into these four storage regions simultaneously.
Each storage section (top and bottom) is connected to a horizontal shift register along its edge. The chip features four output amplifiers, arranged in two pairs: the top storage area feeds into two amplifiers (designated G and H), located on the upper left and right sides of the chip, while the bottom storage area supplies the other two amplifiers (designated E and F), situated on the lower left and right. This configuration effectively divides the sensor into upper and lower halves, each capable of being read out in parallel through two amplifiers, resulting in four-channel simultaneous readout. The entire exposure and readout sequence can be broken down into the following consecutive stages:

\begin{enumerate}

\item \textbf{Integration (Exposure):}Photons accumulate in the image area, generating charge.
\item \textbf{Split Frame Transfer:}After the exposure ends, charge in the image area is rapidly and simultaneously vertically transferred: the top half of the charge is transferred to the top store area and the bottom half of the charge is transferred to the bottom store area.
\item \textbf{Serial Readout Cycle (Example: Top Store Area):}The store area is read out line by line. For each row, the process through the horizontal register and amplifiers includes:
\begin{enumerate}
\item \textbf{Prescan pixel readout:} Before transferring valid image pixels, the horizontal register is clocked through a series of additional cycles. These pixels are read from a prescribed, masked (light-shielded) region outside the image store area (containing no photo generated charge). The output value of these prescan pixels represents the electronic baseline offset (bias level) and read noise of the CCD. They are used in subsequent data processing to establish a precise "zero" reference level for each row.
\item \textbf{Valid image pixel readout:} Afterwards, the horizontal register begins shifting out the valid image charge packets. Pixels from the left half of the row are read out serially through Amplifier G. Simultaneously, pixels from the right half of the row are read out serially through Amplifier H.
\item \textbf{Overscan pixel readout:} After all valid image pixels have been read out, the horizontal register is clocked for additional cycles. First, pixels from the prescribed masked region outside the image storage area are read out, followed by those from the prescribed masked region within the image storage area. The output value of these overscan pixels is used to measure and correct for any charge transfer inefficiency (CTI) issues and clock feedthrough artifacts introduced during the serial readout process, ensuring data integrity.
\end{enumerate}

\item \textbf{Four-Way Synchronization:} While the top store area is reading out one row (including prescan, valid pixels, and overscan) through amplifiers G and H, the bottom store area simultaneously and synchronously performs the identical operation through amplifiers E and F.

\end{enumerate}

\begin{figure}
   \centering
 \includegraphics[width=\textwidth, angle=0, scale=1.0]{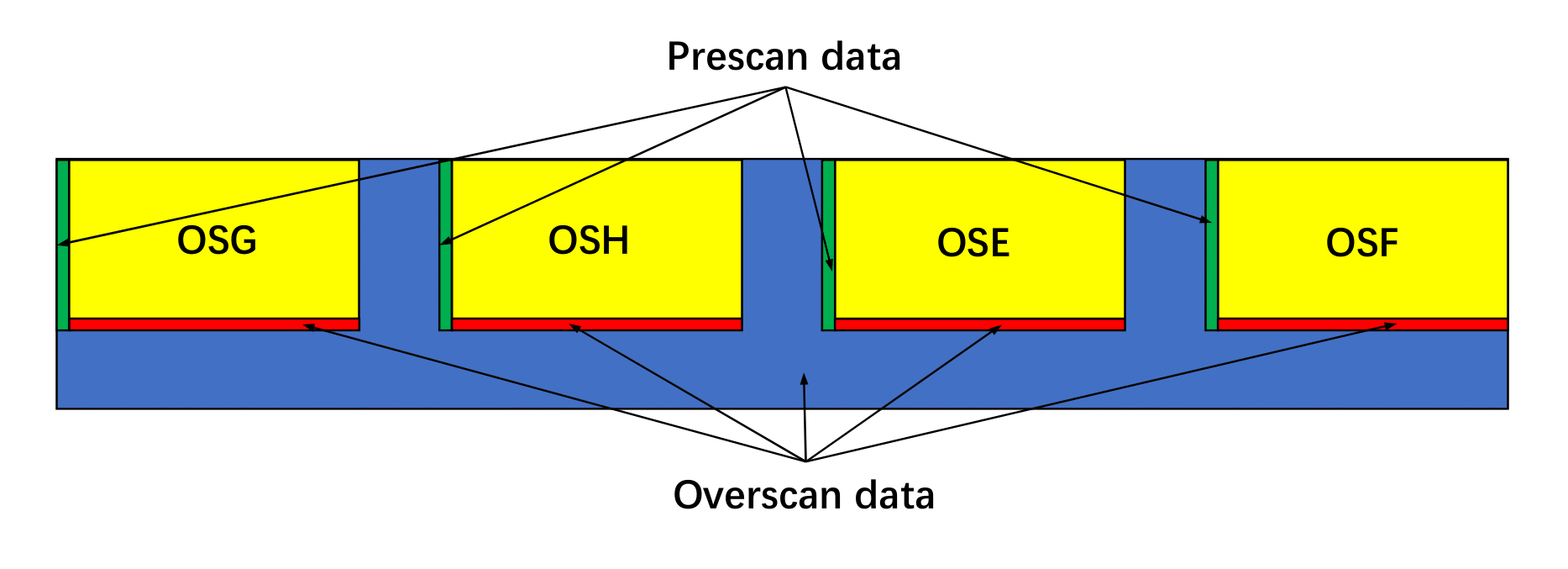}
   \caption{Diagram of the level 0 scientific data format.}
   \label{level-0-image}
\end{figure}

Finally, the processed data from all four output parts (designated OSG, OSH, OSE, and OSF) is reassembled to form a complete, level 0 scientific image, as illustrated in Figure~\ref{level-0-image}. The yellow regions correspond to data from the imaging area. The green areas represent data obtained from the prescan pixel readout of the storage area. The red areas contain data derived from the overscan pixel readout of the imaging area, while the blue areas consist of data originating from the overscan pixel readout of the storage area. 

Notably, while the stored charge is being read out, the photosensitive region continues to integrate signal. During the vertical transfer of charge from the imaging area to the storage region, the ongoing exposure in the light-sensitive area introduces vertical streaking (smearing), particularly along the transfer direction. This results in elevated background gray levels in the form of bright vertical streaks around high-intensity sources such as stars or bright objects, thereby locally increasing the gray values in affected regions. It is evident that frame transfer devices generally exhibit higher frame rates in comparison to full frame devices. A salient benefit of frame transfer devices is their high duty cycle, which signifies that the sensor perpetually collects light. For the detectors CCD230-84 and CCD231-C6, the total frame transfer times are 0.09216 and 0.13824 seconds, respectively. We have incorporated the frame transfer effect into our simulations to evaluate its influence on image gray values, following the exposure and readout sequence principles detailed above. Especially for bright sources, the frame transfer effect cannot be neglected.

\subsection{Other effects}
The pixel size is 15 $\mu$m square and the sensors in the blue and the red channels of the IFS have an image area of 2048 $\times$ 4096 and 3072 $\times$ 6144 pixels, respectively. Each sensor has four outputs for short read-out times. Therefore, each chip consists of four readout channels that are equipped with diverse simulated gains, dark current noise, readout noise, and bias values. Different readout channels have different gain values, and the typical value is 1.5 $e^{-1}/$ADU. The typical value of the dark current noise in each output channel is 0.001 $e^{-1}/sec/pixel$. The typical readout noise with a Gaussian distribution varies from 4.0 to 5.0 $e^{-1}/sec/pixel$ for various channels. For different channels, the typical bias values vary from 450 to 550 ADU. The camera is equipped with a 16-bit depth, and the maximum gray value is 65535. If the charge in a pixel exceeds the saturation level (i.e., the full-well capacity), the pixel becomes saturated and the excess charge will spill over to adjacent pixels. The simulation takes into account bad and hot pixels. The final FITS files containing the two channel image outputs are created, and the data structure is mapped to the CSST Level 0 scientific data format as shown in Figure~\ref{level-0-image}.

It is important to note that certain instrument effects are always present during operation, with no user-configurable parameters available. These inherent effects include frame shift effect, prescan and overscan effect, and A/D conversion processes, and so on. For configurable settings, a comprehensive list of user-adjustable parameters and their descriptions can be found in the reference appendix.

\section{IFS simulation results}
\label{sect:5}
Using the IFS instrument simulation software, simulated calibration images can be produced. The simulated data products comprise standard calibration frames—including dark-field images, flat-field images from a calibration lamp, and wavelength-calibration spectra from a mercury-argon lamp. Furthermore, the simulator generates scientific images for a variety of astronomical sources, with the capability to model specific spectral features such as stellar continua. These results will be presented and discussed in the following section, accompanied by a quantitative evaluation of key instrument performance parameters.

\subsection{Flat calibration simulation results}

\begin{figure}
   \centering
 \includegraphics[width=\textwidth, angle=0, scale=0.8]{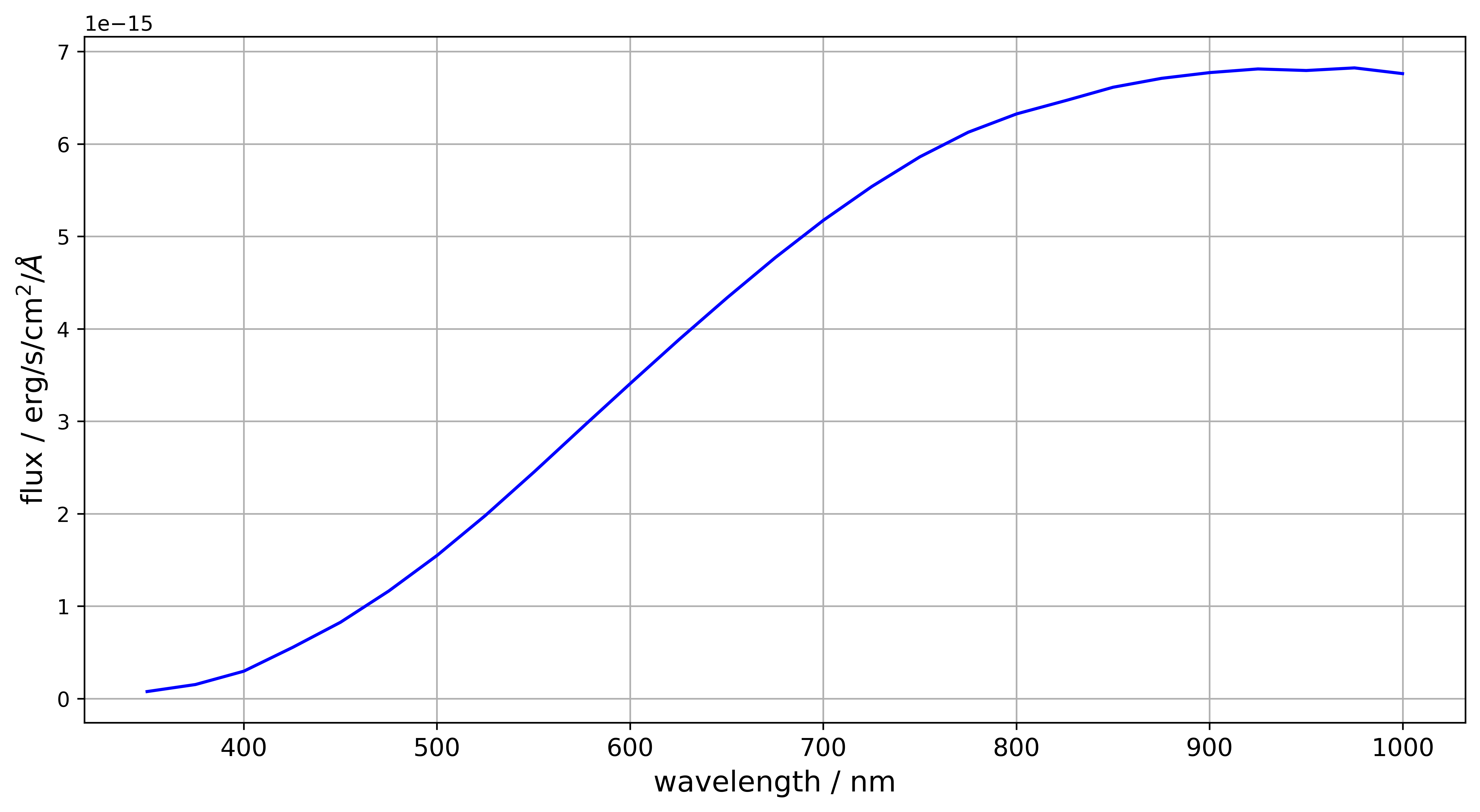}
   \caption{Flux curve of a flat light source.}
   \label{Fig7}
\end{figure}

\begin{figure}
   \centering
 \includegraphics[width=\textwidth, angle=0, scale=1.0]{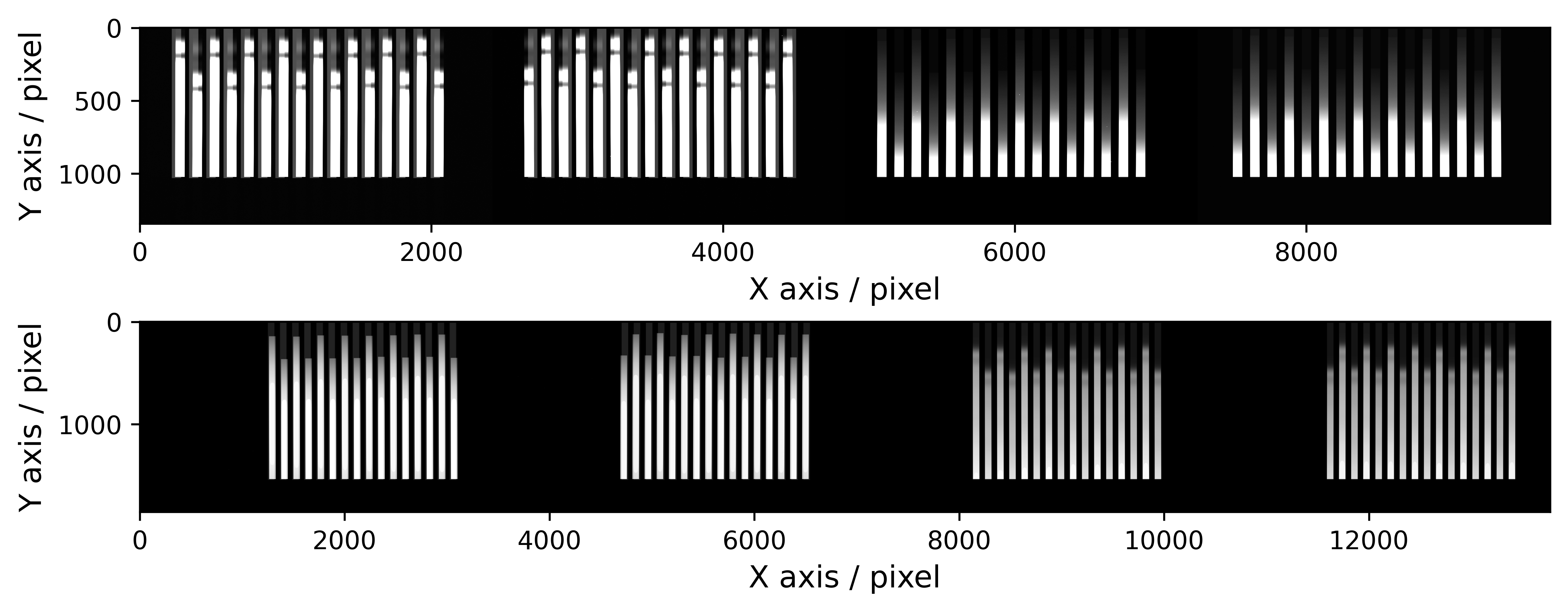}
   \caption{Simulated images of the flat lamp, blue channel image (top), red channel image (bottom).}
   \label{Fig11}
\end{figure}

The IFS is equipped with both a flat spectral calibration lamp and a mercury-argon spectral calibration lamp. Even when the IFS is in orbit, the instrument can be calibrated using the two lamps inside the system. IFS simulation software also needs to simulate the lamp spectral images to achieve instrument calibration, which enables the subsequent data processing pipeline to process the data and support the research of scientific targets. Figure~\ref{Fig7} shows the spectral flux curve of the flat spectral calibration lamp used in the simulation. The flat light intensity profile was generated by simulating the radiation from a 3000 K blackbody. The intensity level used here corresponds to the flat light intensity applied in the simulation. Since the actual intensity of the internal calibration lamp is adjustable, the simulated spectral intensity was intentionally set to a relatively high value to improve the signal-to-noise ratio of the calibration measurements. Figure~\ref{Fig11} shows the spectral images of the flat spectral calibration lamp on the two detectors. These spectral images are used to trace the spectra of each splicer, i.e., to calibrate the position of the spectra.

\subsection{Lamp calibration simulation results}

\begin{figure}
   \centering
 \includegraphics[width=\textwidth, angle=0, scale=1.0]{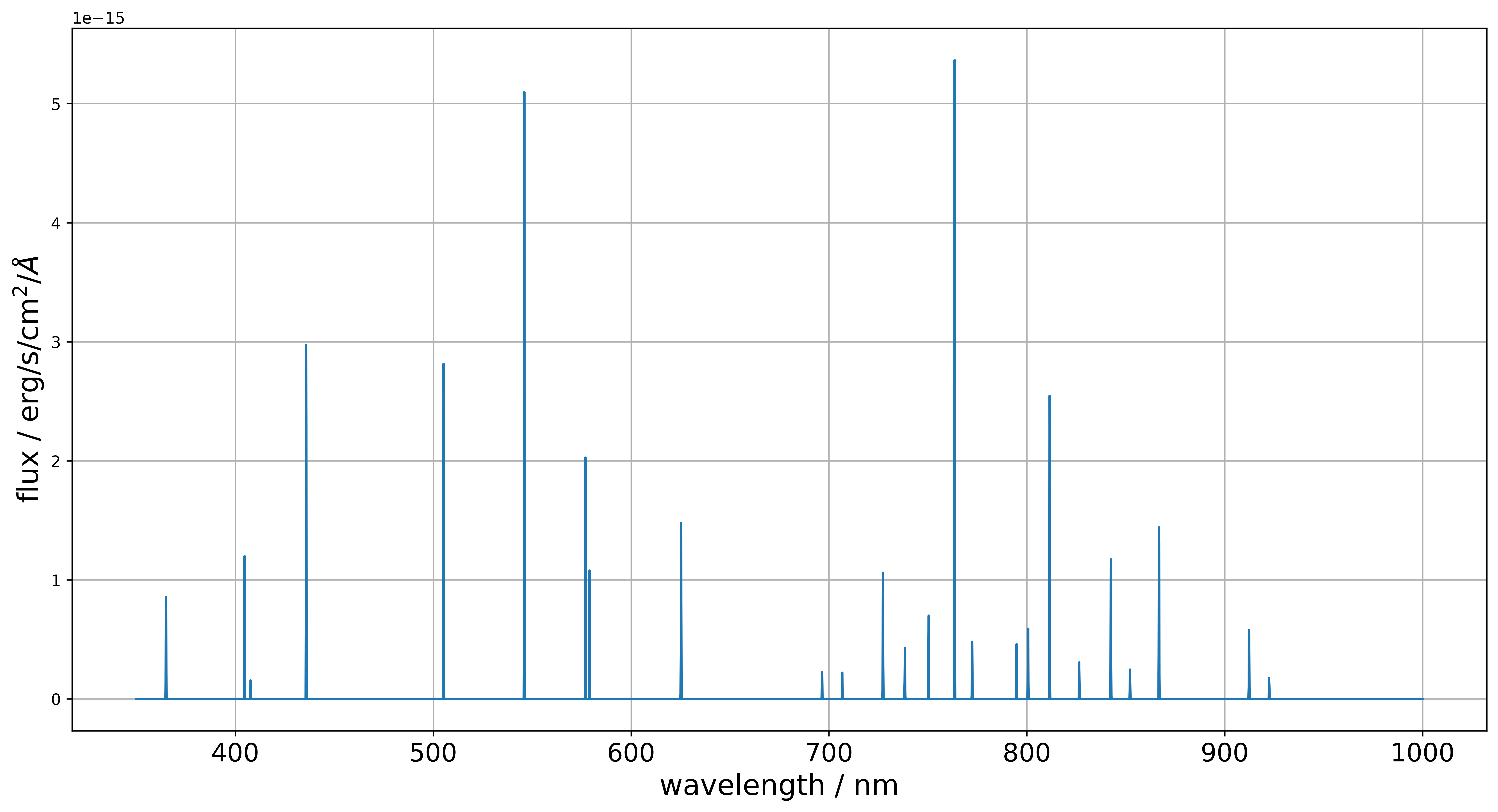}
   \caption{Spectrum flux curve of the mercury-argon lamp.}
   \label{Fig12}
\end{figure}

\begin{figure}
   \centering
 \includegraphics[width=\textwidth, angle=0, scale=1.0]{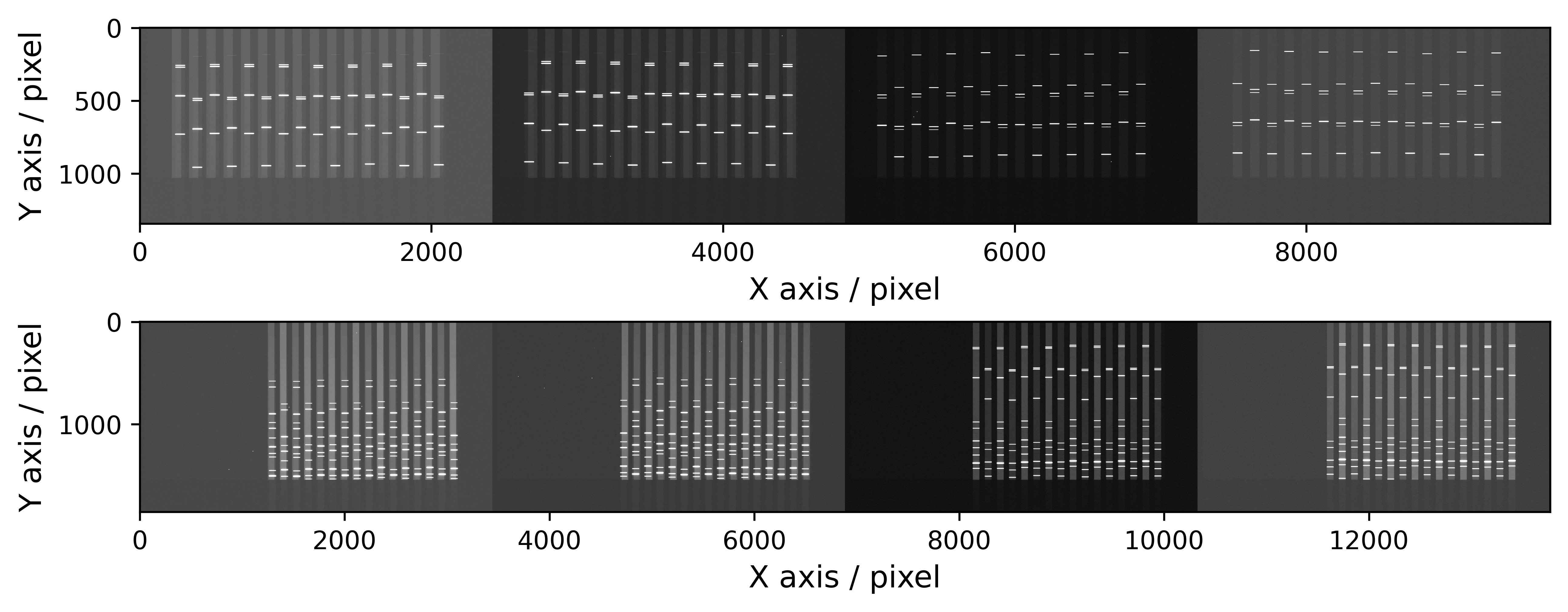}
   \caption{Simulated images of the lamp spectrum of a mercury-argon lamp, blue channel image (top), red channel image (bottom).}
   \label{Fig13}
\end{figure}

Figure~\ref{Fig12} shows the spectral flux curve of the mercury-argon calibration lamp used in the simulation. Figure~\ref{Fig13} shows the spectral images of the mercury-argon calibration lamp on the two detectors. These spectral images are used to establish the wavelength calibration, i.e., to map pixel position to wavelength. 

\subsection{Sky observation simulation results}

\begin{figure}
   \centering
 \includegraphics[width=10cm, angle=0, scale=0.9]{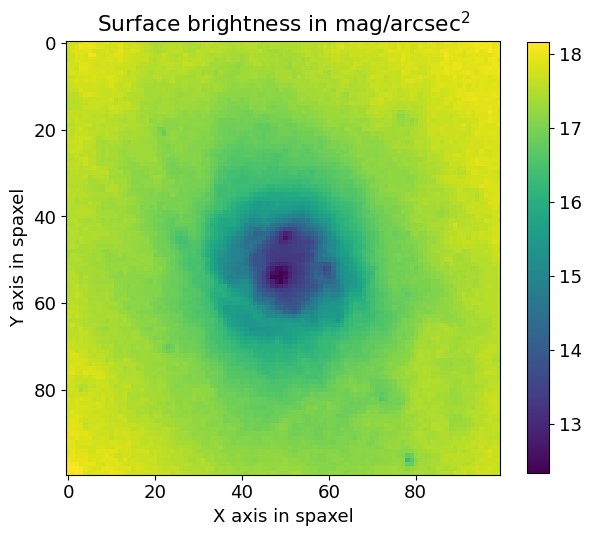}
   \caption{The $g-$band surface brightness of NGC6217.}
   \label{Fig14}
\end{figure}

\begin{figure}
   \centering
 \includegraphics[width=\textwidth, angle=0, scale=1.0]{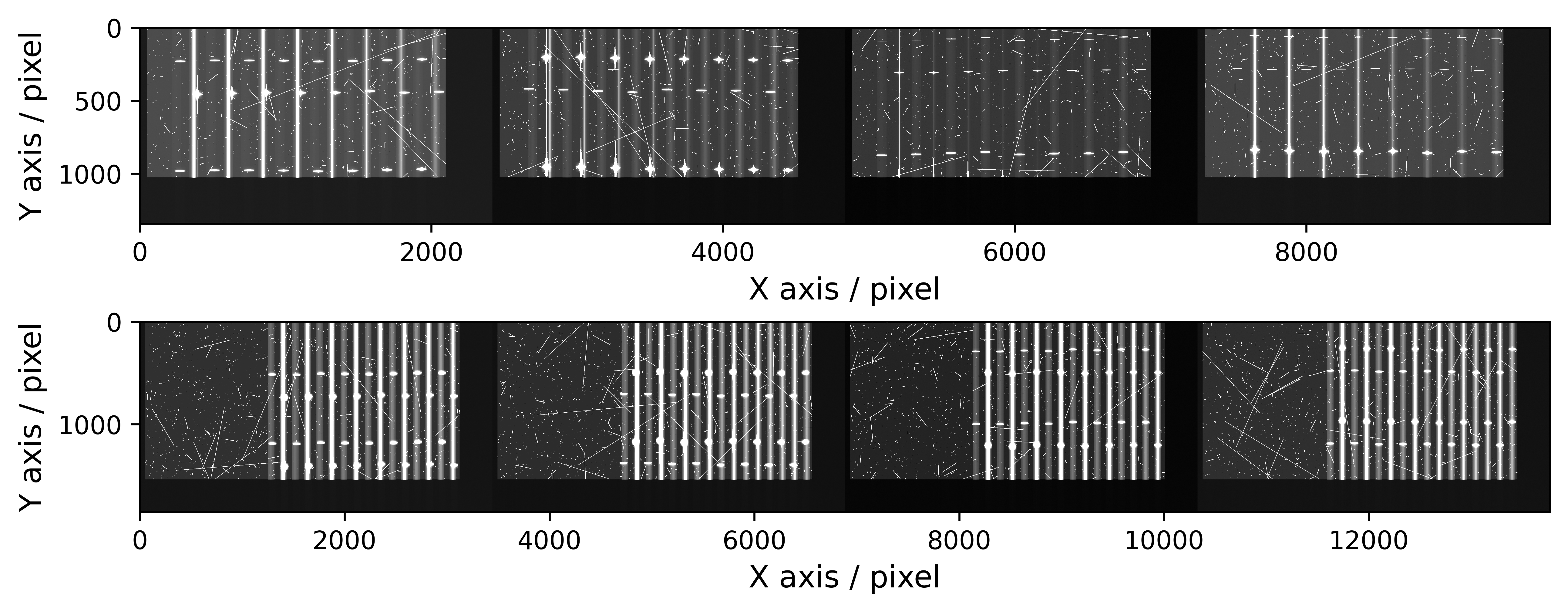}
   \caption{The simulated spectral images of NGC6217, blue channel image (top), red channel image (bottom).}
   \label{Fig15}
\end{figure}

Figure~\ref{Fig14} shows the $g-$band surface brightness of the spiral galaxy NGC 6217 with the pixel scale of 0.1\arcsec. The 3D spectral cube data is generated by the Gehong software. This datacube is used as the input data of the astronomical spectrum of the IFS instrument simulation software, and the final spectral images simulated on the detector are shown in Figure~\ref{Fig15}. After the spectral simulation data is processed by IFS data processing pipeline, the parameters of many observational sources can be obtained, including: gas emission lines, absorption lines, spectral equivalent widths, stellar population parameters such as age, metallicity, extinction, and stellar mass-to-light ratio, etc. Additionally, key performance metrics for the observation, such as the signal-to-noise ratio and exposure time, can also be derived.

\subsection{PSF and LSF measurements}

The PSF and LSF are two core instrumental effects critical to the performance and data interpretation of the IFS. PSF directly determines the spatial resolution of the system -- a narrower PSF means the system can distinguish closer point sources (e.g., resolving two adjacent stars in astronomy). LSF, conversely, quantifies the spectral broadening of a source—such as a line or point source—whose input spectral profile is sufficiently narrow to be approximated as a delta function. The extent of this spectral broadening directly determines the spectral resolution of the system.

To illustrate the reproduction of these instrumental effects, simulated calibration images were generated using a perforated plate. The plate features a uniform array of \SI{5}{\um} diameter holes spaced \SI{100}{\um} apart, positioned on the focal plane of the primary optical system. When illuminated by a calibration lamp, the plate produces thousands of bright spots on the detectors. The horizontal profiles of these spots correspond to the spatial direction, characterizing the PSF, while the vertical profiles align with the dispersion direction, representing the LSF.

\begin{figure}
   \centering
 \includegraphics[width=\textwidth, angle=0, scale=0.9]{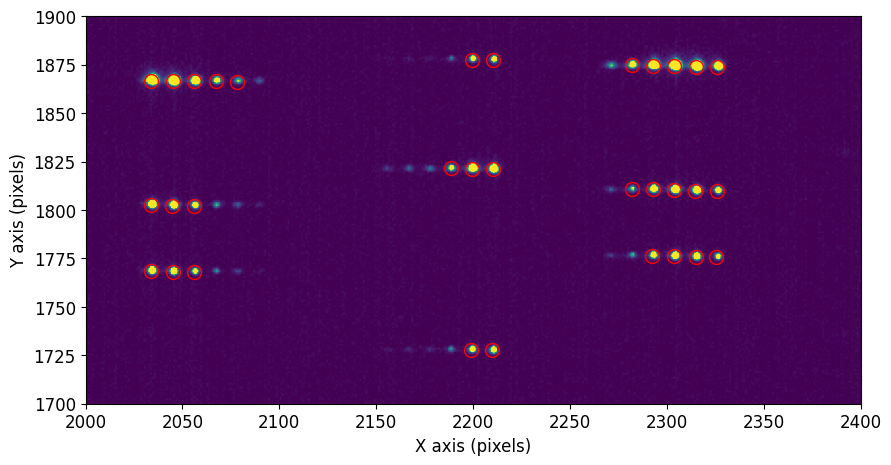}
   \caption{A small segment of the simulated perforated plate illuminated by a calibration lamp. The bright spots with a signal-to-noise ratio (SNR) greater than 10 are marked with red circles.}
   \label{Fig16}
\end{figure}

In Figure~\ref{Fig16}, a small segment of the simulated image in the red channel is displayed, with light spots exhibiting a signal-to-noise ratio (SNR) greater than 10 highlighted by red circles. These selected spots serve as the basis for PSF and LSF measurements. The contour of the bright spot is fitted using a two-dimensional Gaussian function,
\begin{equation}
    f(x, y) = A\exp\left[-\frac{(x - x_0)^2}{2\sigma_{\text{PSF}}^2} + \frac{(y - y_0)^2}{2\sigma_{\text{LSF}}^2}\right]+C,
	\label{eq:gau2D}
\end{equation}
where ($x_0, y_0$) is the light center of the spot, the two sigmas ($\sigma_{\text{PSF}}$ and $\sigma_{\text{LSF}}$) represent the standard deviations in the horizontal and vertical directions, respectively.

\begin{figure}
   \centering
 \includegraphics[width=\textwidth]{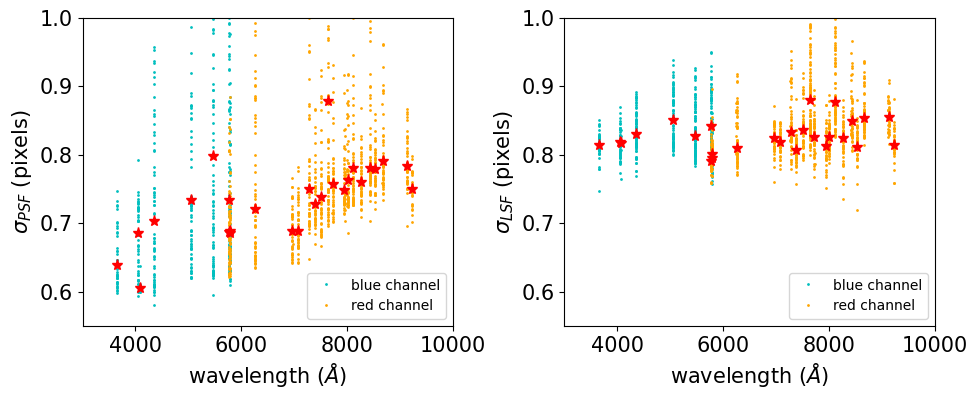}
   \caption{The measured PSF (the left panel) and LSF (the right panel) versus the wavelength of the calibration lamp. The blue dots represent the blue channel, the orange dots represent the red channel, and the red five-pointed stars represent the median values of the measurement results at the same wavelength.}
   \label{Fig17}
\end{figure}

A total of approximately 1500 light spots were selected for two-dimensional Gaussian fitting. The relationships between the measured $\sigma_{\text{PSF}}$ and $\sigma_{\text{LSF}}$ values and the wavelength of the calibration lamp are shown in Figure~\ref{Fig17}. Blue dots correspond to the blue channel, while orange dots represent the red channel. Median values of the measurements at each wavelength are indicated by red five-pointed stars. The measured $\sigma_{\text{PSF}}$ values fall approximately within the range of 0.7–0.8 pixels, consistent with the instrument's design specifications. The measured $\sigma_{\text{LSF}}$ values range from approximately 0.8 to 0.9 pixels, slightly larger than those of the PSF.
This is because, along the spectral dimension, the size of the spot is influenced not only by the spatial PSF, but also extends due to spectral dispersion. As a result, the measured LSF is necessarily broader than the PSF. The instrumental effects demonstrated by the above simulations and analyses are consistent with expectations.

Notably, the performance parameters of the CSST-IFS implemented in the current simulation may deviate from those of the actual system. The PSF and LSF characteristics of the IFS are influenced by several factors that are difficult to quantify individually in practice. These include optical aberrations from the entire telescope and spectrograph system (CSST/IFS), as well as the slit width and alignment inaccuracies. A Monte Carlo approach will be employed in future work to tune the simulation parameters, placing specific emphasis on OPD1, OPD2, and OPD3, thereby enhancing the agreement with experimental data.

As this paper concentrates on the implementation of the IFS instrument simulator, a thorough analysis of influencing factors—such as optical aberrations, slit geometry, and the coherence properties of light—will be reserved for a separate publication. Moreover, due to the inherent complexity of the IFS data and its processing, a detailed presentation of the IFS data reduction pipeline will also be provided in a follow-up paper. Furthermore, as our simulations yield raw CCD image data that do not permit direct estimation of the signal-to-noise ratio—instead necessitating sophisticated reduction software for processing—this workflow presents a significant accessibility barrier for typical scientific users. To address this, we have developed the Exposure Time Calculator (ETC) simulator software, which is now publicly available online (\url{https://nadc.china-vo.org/csst-bp/etc-ifs}). As part of the CSST-IFS article series, the ETC will be described in detail in a dedicated forthcoming paper. There, we will present comprehensive demonstrations of signal-to-noise ratio estimations for various types of target sources, along with practical recommendations for observational planning.

\section{Conclusions}
\label{sect:conclusion}
We have developed a comprehensive end-to-end simulation framework for the CSST-IFS, designed to generate realistic synthetic data by modeling the full optical path and incorporating major instrumental and environmental effects. Key components—including the CSST primary telescope, image slicers, pupil mirror and slit array, and Offner spectrometers—have been accurately represented, with wavefront errors integrated into wavelength-dependent PSF calculations.

The simulation accounts for critical effects such as optical diffraction, grating dispersion (including sub-pixel shifts), system throughput, detector characteristics (CTI, PRNU, nonlinearity, frame transfer smearing), and environmental factors (cosmic rays, sky background, and Doppler effect). This enables the production of calibrated flat-field and wavelength calibration images, as well as scientific observations such as that of galaxy NGC 6217. The series of simulated calibration reference files and scientific targets is crucial for supporting the subsequent development of the IFS data processing software (Yin et al., in preparation)


Prior analysis and processing of simulated spectral data confirm that the existing simulation software faithfully reproduces the key design specifications of the IFS hardware. Consequently, this work offers a robust tool for observation planning, pipeline validation, and feasibility assessment. To enhance the realism of the simulations, we have implemented a three-phase refinement strategy: (1) integrating laboratory-measured parameters into future versions of the software; (2) performing comprehensive pre-launch calibration; and (3) refining the models using post-launch on-orbit data. This iterative process ensures continuous improvement of the simulation framework, enabling increasingly accurate support for mission preparation and scientific analysis as further measurements are obtained.

\begin{acknowledgements}
This work was funded by the National Natural Science Foundation of China (NSFC) under No.11080922, No.11873078 and No.12573115. J.Y. acknowledges support from the Natural Science Foundation of Shanghai (Project Number: 22ZR1473000) and the Program of Shanghai Academic Research Leader (No. 22XD1404200)
\end{acknowledgements}

\bibliographystyle{raa}
\bibliography{bibtex}

\label{lastpage}
\appendix
\section{Instrument effect switch}
The code for the CSST-IFS instrument simulation software can be downloaded at the following URL: \url{https://csst-tb.bao.ac.cn/code/csst-sims/csst_ifs_sim}. Some instrument effects are always on in the simulation and cannot be changed by the user. The instrument effect switches that can be set by the user are as follows.

\# apply multiplicative flat field (to emulate pixel-to-pixel non-uniformity) or not

$\bullet$ flatfieldM = yes

\# add CCD dark current or not

$\bullet$ darknoise = yes

\# add skylight background and stray light or not

$\bullet$ sky\_noise = yes

\# apply cosmetic defects to CCD or not

$\bullet$ cosmetics = yes

\# apply radiation damage effect or not

$\bullet$ radiationDamage = yes

\# add cosmic rays or not  

$\bullet$ cosmicRays = yes

\# apply bleeding effect or not

$\bullet$ bleeding = yes

\# apply non-linearity effect or not

$\bullet$ nonlinearity = yes

\# apply readout noise or not

$\bullet$ readoutnoise = yes

\# save cosmicrays map or not

$\bullet$ save\_cosmicrays = no

\end{document}